\documentclass[structabstract]{aa}
\usepackage{graphicx}
\usepackage{natbib}
\usepackage{amssymb}
\usepackage{longtable}
\usepackage{lscape}
\usepackage{url}
\usepackage{soul} %%%\st{Hellow world}

\usepackage{color}

\usepackage{txfonts}
\begin{document}
   \title{A new look at the long-period eclipsing binary V383\,Sco
\thanks{Based on data from the All Sky Automated Survey (ASAS-3) conducted
by the Warsaw University Observatory (Poland), at the Las Campanas
Observatory, on observations collected at the La Silla Paranal Observatory,
ESO (Chile), with the HARPS spectrograph at the 3.6\,m telescope (ESO run
084.D-0591(A)), and on a low-resolution spectrum obtained at the South
African Astronomical Observatory (SAAO) with the Grating Spectrograph at the
1.9\,m Radcliffe telescope.  Data from the Appendix (Tables \ref{ASAS_V.dat}
-- \ref{ASAS_V-I_int.dat}) are only available in electronic form at the CDS
via anonymous ftp at \protect\url{cdsarc.u-strasbg.fr} or
%(130.79.128.5) %% no point putting IP number IMHO --boud 
via \protect\url{http://cdsweb.u-strasbg.fr/cgi-bin/qcat?J/A+A/}}}

%   \subtitle{}

\author{Cezary Ga{\l}an\inst{1}
\and Toma~Tomov\inst{1}
\and Taichi~Kato\inst{2}
\and Grzegorz~Pojma\'nski\inst{3}
\and Dorota~M.~Szczygie{\l}\inst{3,4}
\and Bogumi{\l}~Pilecki\inst{3,5}
\and Dariusz~Graczyk\inst{5}
\and Mariusz~Gromadzki\inst{6}
\and Maciej~Miko{\l}ajewski\inst{1}
\and Wolfgang~Gieren\inst{5}
\and Andrzej~Strobel\inst{1}
\and Boudewijn~F.~Roukema\inst{1}
	}

%\offprints{Cezary Ga{\l}an (cgalan@astri.uni.torun.pl)}

\institute{
Toru\'n Centre for Astronomy, Nicolaus Copernicus University, ul. Gagarina 11, 87-100 Toru\'n, Poland\\
\email{cgalan@astri.uni.torun.pl}
\and %2
Department of Astronomy, Faculty of Science, Kyoto University, Sakyo-ku, Kyoto 606-8502
\and %3
Warsaw University Astronomical Observatory, Al. Ujazdowskie 4, 00-478 Warszawa, Poland
\and %4
Department of Astronomy, The Ohio State University, 140 W 18th Avenue, Columbus, OH 43210, USA
\and %5
Universidad de Concepci\'on, Departamento de Astronomia, Casilla 160-C, Concepci\'on, Chile
\and %6
Departamento de F{\'i}sica y Astronom{\'i}a, Centro de  Astrof{\'i}sica de Valpara{\'i}so, Universidad de Valpara{\'i}so, Av. Gran Breta na1111, Playa Ancha, Casilla 5030, Chile
}

   \date{Received xxxxx xx, xxxx; accepted xxxxx xx, xxxx}

  \abstract
   {The~system V383\,Sco was~discovered to~be an~eclipsing binary star
at the~beginning of the~twentieth century.  This system has one
of the~longest orbital periods known (13.5\,yr) and was~initially classified
as~a~$\zeta$Aur-type eclipsing variable.  It~was then forgotten about for
decades, with no~progress made in~understanding it.}
   {This study provides a~detailed look at~the~system V383\,Sco, using new
data obtained before, during and~after the~last eclipse, which occurred
in~2007/8.  There was a~suspicion that this system could be~similar
to~eclipsing systems with extensive dusty disks like EE\,Cep and
$\varepsilon$\,Aur.  This and~other, alternative hypotheses are considered
here.}
   {The All~Sky~Automated~Survey (ASAS-3) $V$~and~$I$~light curves have been
used to~examine apparent magnitude and~colour changes.  Low- and
high-resolution spectra have been obtained and~used for~spectral
classification, to~analyse spectral line profiles, as~well as~to~determine
the~reddening, radial velocities and the~distance to the~system. 
The~spectral energy distribution (SED) was~analysed using all~available
photometric and~spectroscopic data.  Using our own~original numerical code,
we~performed a~very simplified model of the~eclipse, taking into account
the~pulsations of~one of the~components.}
   {The~low-resolution spectrum shows apparent traces of~molecular bands,
characteristic of an~M-type supergiant.  The presence of~this star
in~the~system is~confirmed by~the~SED, by~a~strong dependence of the~eclipse
depth on~the~photometric bands, and by~the~nature of~pulsational changes. 
The presence of~a~very low excitation nebula around the~system has~been
inferred from [\ion{O}{i}] 6300\,\AA\ emission in~the~high-resolution
spectrum.  Analysis of~the~radial velocities, reddening, and
period-luminosity relation for~Mira-type stars imply a~distance
to~the~V383\,Sco system of~8.4\,$\pm$\,0.6\,kpc.  The~distance to~the~nearby
V381\,Sco is 6.4\,$\pm$\,0.8\,kpc.  The~very different and~oppositely
directed radial velocities of~these two~systems ($89.8$ km\,s$^{-1}$
vs~$-178.8$ km\,s$^{-1}$) seem to~be in~agreement with a~bulge/bar kinematic
model of the~Galactic centre and~inconsistent with purely circular motion.}
   {We~have~found strong evidence for the~presence of~a~pulsating M-type
supergiant in the~V383\,Sco system.  This supergiant periodically obscures
the~much more luminous F0\,I-type star, causing the~deep (possibly total)
eclipses which vary in~duration and~shape.}

   \authorrunning {C. Ga{\l}an et al.}

   \titlerunning {A new look at the long-period eclipsing binary V383\,Sco}


   \maketitle
%
%________________________________________________________________

   \section{Introduction}

The star system V383\,Sco (HV 7021) was discovered to be an eclipsing binary
at the beginning of the twentieth century during a~photographic study of
variable stars in a~field of the Milky Way near the Galactic centre.  As
a~partial result of those studies, Henrietta \citet{Swo1936} presented
photometric data containing observations of three eclipses whose minima
occurred in 1901, 1914, and 1928.  The very long orbital period,
approximately $4900^{\rm d}$ (13.5 yr), is one of the longest known among
eclipsing binaries.  Its light curve has a~wing-like shape at the beginning
and at the end of the eclipses, which is characterized by slower photometric
changes.  Henrietta Swope considered it to be an~effect of atmospheric
absorption and suggested that V383\,Sco is similar to $\zeta$\,Aurigae-type
stars.  Later, the spectral type of the primary was estimated as F0Ia
\citep{Pop1948}.  V383\,Sco was then neglected for many years and
\citet{OCon1951} barely mentioned it during his analysis of the~light curve
of the V644\,Cen system.  Currently V383\,Sco is a~very poorly--studied
system.

During the past decade the ASAS-3 survey has monitored V383\,Sco in the
standard $V$ and $I$ photometric bands.  These observations cover the last
eclipse, from 2007 to 2008 (see Figs~\ref{ASAS3.VI}~and~\ref{ASAS3.ecl.VI}). 
While studying the $V$ light curve, one of us pointed out the asymmetry of
this eclipse \citep{Kat2008} and its similarity to the asymmetry of eclipses
observed in the EE\,Cep system \citep[see][]{Gal2012}.  Inspired by this
discovery, we began the~study of V383\,Sco to verify the possibility that
the eclipses could be caused by a~dusty disk, as in the unique EE\,Cep and
$\varepsilon$\,Aur systems.  Here, we present an analysis based on the
ASAS-3 survey $V$- and $I$-band photometry, low and high-resolution spectra,
and all available visual, near- and far-infrared photometric data.

%
%______________________________________________________________

\section{Observations and data reduction}

\begin{figure}
  \resizebox{\hsize}{!}{\includegraphics{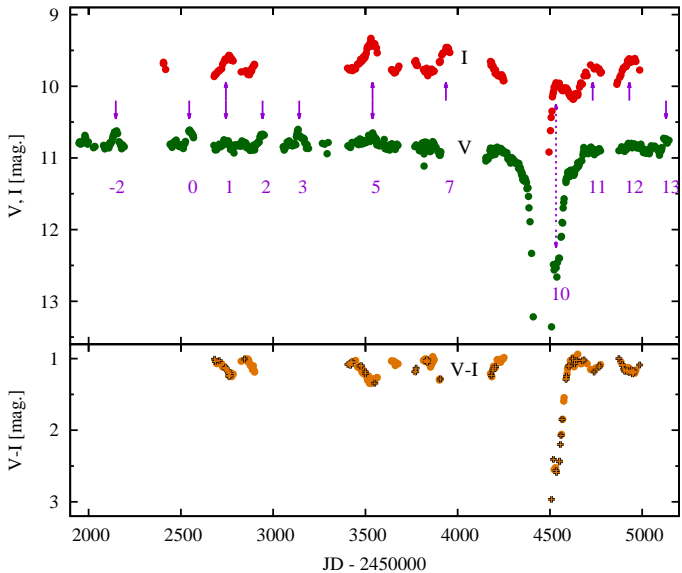}}
  \caption{Top: $V$- and $I$-band light curves from ASAS-3.  Bottom: the
$V-I$ colour index.  Crosses denote colour indices $V-I$ obtained by
interpolation in the cases when $V$ and $I$ measurements were not
simultaneous, i.e.  not taken during the same nights.  Arrows and numbers
show the moments of the pulsation maxima visible in $V$- and $I$-band light
curves, used for timing analysis of pulsations (see
Sect.~\ref{sec.pulsations}).  The dotted arrow shows a postdiction for the
pulsation maximum at epoch $E_{\mathrm p} = 10$ (corresponding to a maximum
that occurred during the most recent eclipse --- epoch $E = 8$) according to
the ephemeris from Eq.~\ref{pulslinefem}.}
  \label{ASAS3.VI}
\end{figure}

\begin{figure}
  \resizebox{\hsize}{!}{\includegraphics{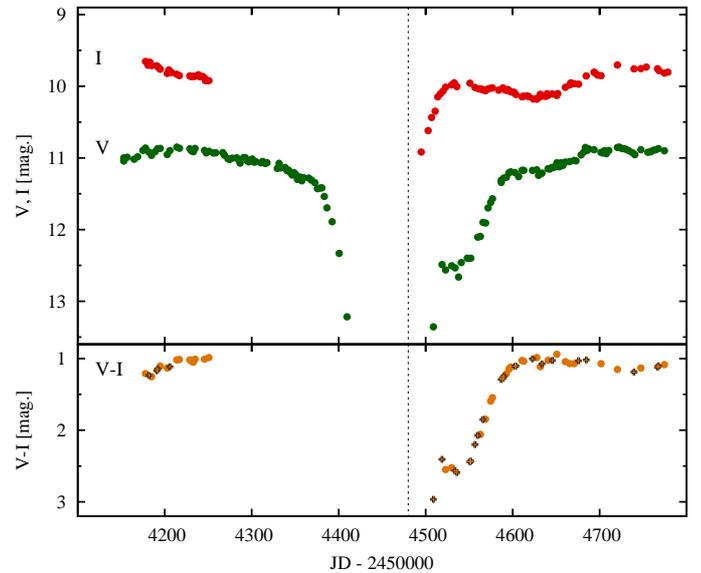}}
  \caption{Expanded view of the 2007/8 eclipse of V383\,Sco. The symbols are
the same as in Fig.~\protect\ref{ASAS3.VI}.  The vertical dashed line marks
the mid-eclipse moment$^2$ (Table~\ref{ocl.t}).}
  \label{ASAS3.ecl.VI}
\end{figure}

The ASAS-3 survey \citep[see][]{Poj2004} has monitored V383\,Sco from the
Las Campanas Observatory (Chile) with two standard filters since 6~Feb.~2001
(JD\,2541947) in $V$ and since 5~Sept.~2002 (JD\,2452404) in $I$.  The data
were extracted from the database in April 2010 and processed using the
pipeline described by \citet{Poj1997}.  We selected data obtained with
diaphragm numbers 2 and 0 in the cases of bands $V$ and $I$, respectively,
which had the smallest statistical errors.  The resulting $V$ and $I$-band
light curves are presented in Fig.~\ref{ASAS3.VI} with the colour index
$V-I$ which has been calculated for measurements made on the same nights
(marked with circles) and by interpolation of close but non-simultaneous
measurements (marked with crosses).  These are available as
Tables~\ref{ASAS_V.dat}--\ref{ASAS_V-I_int.dat} in our online Appendix.

High-resolution spectra (R=84000) of V383\,Sco and the nearby long-period
($P = 6545^{\mathrm d}$) eclipsing binary V381\,Sco in the spectral range
3800--6900\,\AA\AA\/ were obtained at La Silla with the High Accuracy Radial
velocity Planet Searcher (HARPS) spectrograph using the EGGS mode on two
consecutive nights in mid-October 2009.  The spectra were extracted and the
wavelength calibrated using the HARPS pipeline.  Additionally,
a~low-resolution spectrum was acquired at the end of October 2009 with the
Grating Spectrograph with a SITe (Scientific Imaging Technologies, INC.) CCD
mounted at the 1.9\,m Radcliffe telescope at the South African Astronomical
Observatory (SAAO).  Grating number\,7 with 300 lines\,mm$^{-1}$ and a~slit
width of $1\farcs5$ was used.  To calibrate the spectrum, the
spectrophotometric standard stars LTT\,2415, LTT\,7987, and LTT\,9239 were
used.  All the data reduction and calibrations were carried out with
standard IRAF\footnote{IRAF is distributed by the National Optical Astronomy
Observatory, which is operated by the Association of Universities for
Research in Astronomy (AURA) under co-operative agreement with the National
Science Foundation.} procedures.  The extracted and flux-calibrated spectrum
of V383\,Sco covers the range $\sim$3800--7700\,{\AA} with a~resolving power
of $R \simeq 1000$.  The journal of our spectroscopic observations is given
in Table~\ref{jur}.

\begin{table}
\caption{Journal of spectroscopic observations.}
\label{jur}
\centering 
\begin{tabular}{llllll}
\hline\hline
Star     & Date       & HJD (mid)    & Exposure          & Observat.\\
\hline
V383~Sco & 14.10.2009 & 2455119.4965 &  900$^\mathrm{s}$ & ESO      \\
V381~Sco & 15.10.2009 & 2455119.5115 & 1100$^\mathrm{s}$ & ESO      \\
V383~Sco & 16.10.2009 & 2455120.5610 & 1100$^\mathrm{s}$ & ESO      \\
V383~Sco & 31.10.2009 & 2455136.2602 & 1800$^\mathrm{s}$ & SAAO     \\
\hline
\end{tabular}
\end{table}

%
%______________________________________________________________

\section{Results and discussion}

\subsection{The eclipses and orbital period}

The $V$- and $I$-band light curves and $V-I$ colour index of the last
eclipse of V383\,Sco (in 2007/8) are shown in
Figs.~\ref{ASAS3.VI}~and~\ref{ASAS3.ecl.VI}.  There is a strong dependence
of the eclipse depth on the~photometric band.  The mid-eclipse on
14~Jan~2008 (JD\,2454480) occurred about $170^{\rm d}$ earlier than
predicted from the ephemeris

\begin{figure}
  \resizebox{\hsize}{!}{\includegraphics{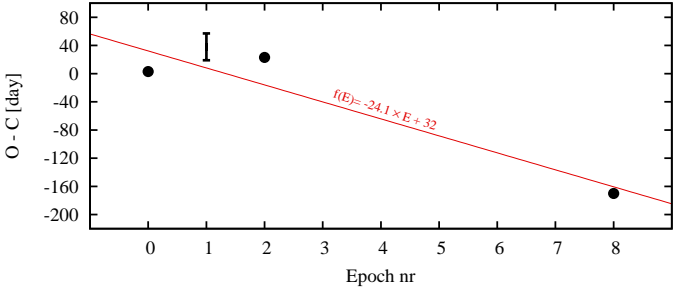}}
  \caption{$O-C$ diagram with a linear (solid line) fit to the residuals. 
The moment of minimum at epoch 1 (vertical bar) is an estimated figure
because adequate data were not available.}
  \label{ocl}
\end{figure}

\begin{center}
\begin{equation}
JD_{\rm midecl} = 2415450 + 4900 \times E  \label{archlinefem}
\end{equation}
\end{center}

\noindent which was constructed with the period and the moment of minimum
from \citet{Swo1936}.  We have used the archival photometry of three minima
observed before 1930 by \citet{Swo1936} (where the upper estimates and
uncertain points were excluded from the analysis) and the ASAS-3 data
obtained in the $V$ band for timing analyses.  Table~\ref{ocl.t} lists
approximate times of mid-eclipses, estimated using a~slightly simplified
variant of the \citet{Kwe1956} method\footnote{The times of mid-eclipse
    were estimated as follows: For the eclipse at
    $E=8$, 
    $T_{\rm 8d}$ is the moment on the descending branch at which the
    brightness has dropped
    $1.^{\rm m}0$ below the out-of-eclipse mean, and 
    for eclipses at $E= 0, 2$ and 8,
    $T_{\rm Ea}$ is
    the moment on the ascending branch at which the brightness reaches
    $1.^{\rm m}0$ below the out-of-eclipse mean.  The 
    mid-eclipse was defined as 
    $T_{\rm Em} = T_{\rm Ea} - (T_{\rm 8a} - T_{\rm 8d})/2$.}.  
The~residuals ($O-C$, observations minus calculations) for the times of
minima were calculated using the linear ephemeris in Eq.~\ref{archlinefem},
are listed in Table~\ref{ocl.t}, and are marked in Fig.~\ref{ocl}.  The best
linear fit to the residuals at epochs 0, 2, and 8, gives the new ephemeris

\begin{center}
\begin{equation}
JD_{\rm midecl} = 2415482 (\pm41) + 4875.9 (\pm 8.5) \times E. \label{newlinefem}
\end{equation}
\end{center}

\begin{figure}
  \resizebox{\hsize}{!}{\includegraphics{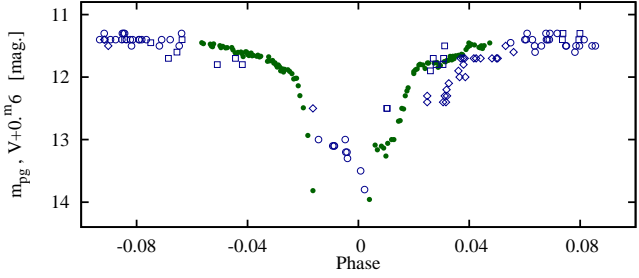}}
  \caption{Light curves of four eclipses phased with a~constant period
according to the new linear ephemeris (Eq.~\ref{newlinefem}).  The open
symbols denote archival photographic photometric data (squares: $E = 0$,
circles: $E = 1$, diamonds: $E = 2$).  The $V$-band ASAS data (filled
circles: $E = 8$) have been shifted arbitrarily by $+0.6$ for facilitate
comparison with the light curves from the different photometric systems.}
  \label{LClp}
\end{figure}

\begin{table}
\caption{Moments of minima and the $O-C$ residuals estimated according to
the linear ephemeris $JD_{\rm midecl} = 2415450 + 4900 \times E $.}
\label{ocl.t}
\centering			% used for centering table
\begin{tabular}{lcc}%\l}
\hline\hline
Epoch       &     $JD_{\rm midecl}$=($T_{\rm Em}$)      &  $O-C$ [day]\\%\ & Ref.\\
\hline
0           &       2415453                 & ~~~3  \\%\ & Swope 1936\\           %15544-91
1$^{\star}$ &       2420369--2420407        & 19--57\\%\ & Swope 1936\\
2           &       2425273                 & ~~23  \\%\ & Swope 1936\\           %25364-91
8           &       2454480                 & -170  \\%\ & ASAS III\\           %(54389+54571)/2 91=(54571-54389)/2
\hline
\end{tabular}
\begin{list}{}{}
\item[$^{\star}$] not used for the ephemeris fitting
\end{list}
\end{table}

It is not possible, however, to phase the light curve correctly
(Fig.~\ref{LClp}) with this new ephemeris.  Regardless of which linear
ephemeris is used, the eclipses are always shifted with respect to each
other by up to several weeks.  This makes the V383\,Sco case reminiscent of
the VVCep system, where the 1997 eclipse occurred by about 1\% of an orbital
period later than predicted \citep{Gra1999}.  The reason for such large
differences in eclipse contact moments is not clear, but could be due to
changes in the orbital parameters.  However, in the case of V383\,Sco it
might be appropriate to consider another possibility.  The changes in the
eclipse contact moments could be caused by variations in the radius of the
eclipsing component caused by pulsations which could explain the observed
changes in the durations of the eclipses -- the eclipse at epoch $E = 2$
clearly lasted significantly longer than the most recent one.

In the observational data collected so far, no traces of an eclipse of the
secondary, cool component, have been detected.  There is no way to predict
the phase of the secondary eclipse, because through the entire orbital
period the spectroscopic observations needed to infer the radial velocity
curves are almost completely lacking.

%
%______________________________________________________________

\subsection{Pulsations} \label{sec.pulsations}

\begin{figure*}
  \centering
    \includegraphics[width=17cm]{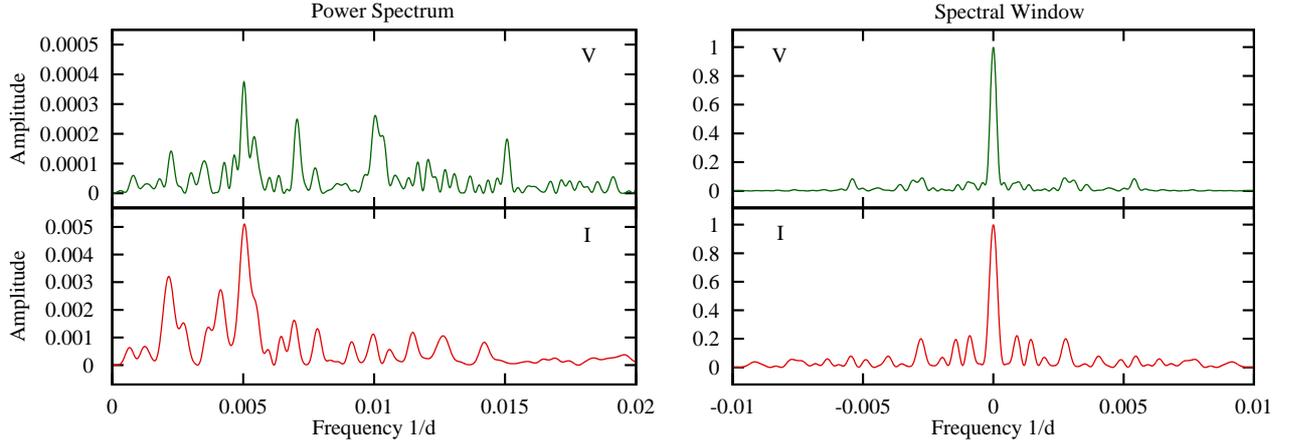}
  \caption{Power spectra obtained using Fourier analysis on out-of-eclipse
    $V$-band (top) and $I$-band (bottom) data (see
Sect.~\protect\ref{sec.pulsations}).}
  \label{Periods}
\end{figure*}

\begin{figure}
  \resizebox{\hsize}{!}{\includegraphics{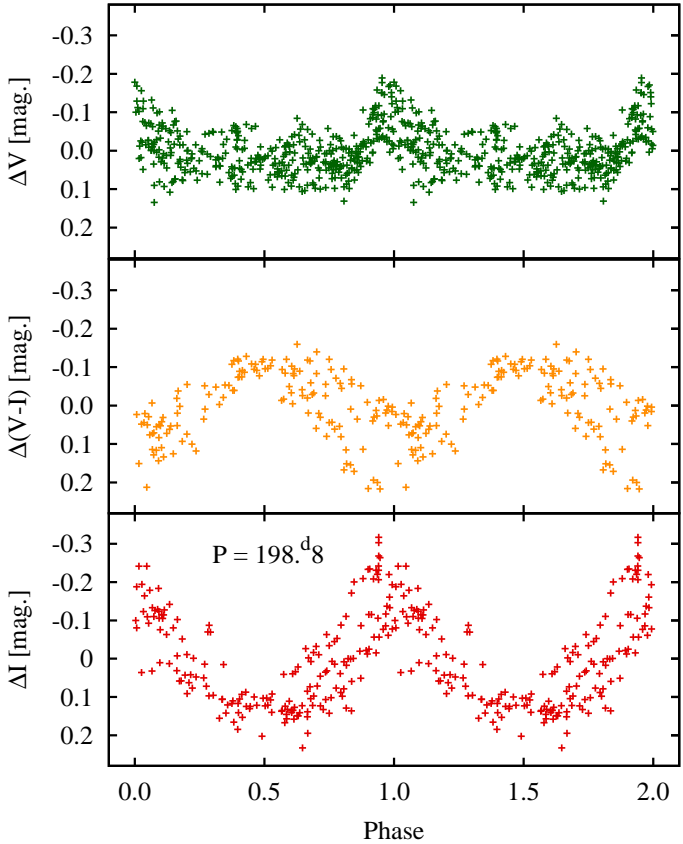}}
  \caption{ASAS-3 $V$-band (top), $V-I$ (middle), and $I$-band (bottom)
light curves outside of the eclipse, phased with period $P_{\mathrm {pul}} =
198$~\fd$8$.  The small amplitude of the observed $V$ variation and its
strongly non-sinusoidal shape with a wide minimum are related to the
presence in the system of the second component which strongly dominates in
this wavelength range (see section \ref{simplemodel}).}
  \label{Phased.outofecl.VI.1p}
\end{figure}

The ASAS-3 photometric data outside of eclipses show apparent variations
which seem to be connected with stellar pulsations.  To investigate the
amplitude and periodicity of these variations, we studied the out-of-eclipse
data in the time intervals JD\,2451947--JD\,2454216 and
JD\,2454683--JD\,2455145.  After eliminating trends from the ASAS $V$, $I$
and $V-I$ data (by fitting and subtraction of a~linear function and/or
a~second order polynomial), a fast Fourier transform was used to search for
possible periods of variation.

\begin{table}
\caption{Frequencies and periods obtained with Fourier analysis of the
out-of-eclipse $V$-band light curve.  The errors $\sigma$ were estimated
using the HWHM$^{\star}$ of peaks in the power spectrum.}
\label{FrAmPe}
\centering			% used for centering table
\begin{tabular}{llll}
\hline\hline
Frequency [1/d]  & Amplitude & Period [d]& $\sigma$ [d]\\
\hline
0.00503    &     0.000375   &      198.8 & 4.8  \\
0.01004    &     0.000262   &      ~99.6 & 1.5  \\
0.01507    &     0.000182   &      ~66.36& 0.49 \\
\hline
\end{tabular}
\begin{list}{}{}
\item[$^{\star}$] Half width at half maximum.
\end{list}
\end{table}

The power spectra obtained using the $V$-band data are presented in
Figure~\ref{Periods} (top) and the frequencies and the corresponding periods
are given in Table \ref{FrAmPe}.  The dominant frequency is the lowest
strong peak at $\sim 0.00505$/d, i.e.  a pulsation period of $\sim198$~\fd0. 
A similar peak dominates in the $I$ (Fig.~\ref{Periods} (bottom)) and $V-I$
data, but less accurate by a factor of about two, while the peaks at greater
frequencies (corresponding to shorter periods) are not visible.  We
therefore suggest that the period $P_{\mathrm {pul}} = 198$~\fd$8 \pm 4.8$
is a~reliable value of a~pulsation period for one of the components in this
system.  The $V$ and $I$-band light curves and variations of the $V-I$
colour index after phasing with this pulsation period are shown in
Figure~\ref{Phased.outofecl.VI.1p}.  The amplitude of variations observed in
the redder $I$ band is much greater than in the bluer $V$ band.

Pulsational maxima $E_{\mathrm p}$ from the $V$- and $I$-band light curves,
as shown in Fig.~\ref{ASAS3.VI}, were used for a second timing analysis. 
The moments of maxima are shown in Table~\ref{tocpul} with their
corresponding $O-C$ residuals, which were calculated using the following
ephemeris

\begin{center}
\begin{equation}
JD_{\rm maxPul} = 2452547.5 + 198.8 \times E_{\mathrm p}. \label{pulsstartefem}
\end{equation}
\end{center}

\begin{figure}
  \resizebox{\hsize}{!}{\includegraphics{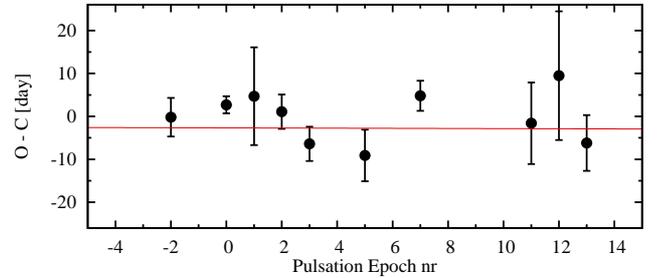}}
  \caption{$O-C$ diagram for the pulsation maxima with a~linear (solid line)
fit to residuals.}
  \label{focpul}
\end{figure}

\noindent The pulsation zero epoch ($E_{\mathrm p} = 0$) was adopted
arbitrarily at JD\,2452547.5.  The best linear fit (Fig.~\ref{focpul}) gives
the ephemeris for pulsation maxima

\begin{center}
\begin{equation}
JD_{\rm maxPul} = 2452544.8 (\pm 2.9) + 198.8 (\pm 0.4) \times E_{\mathrm p}. \label{pulslinefem}
\end{equation}
\end{center}

\begin{table}
\caption{Pulsational maxima and the $O-C$ residuals estimated according to
the initial ephemeris in Eq.~\ref{pulsstartefem}.}
\label{tocpul}
\centering			% used for centering table
\begin{tabular}{rrcc}
\hline\hline
$E_{\mathrm p}$ & JD-2400000  & uncertainty [d] & O-C [d]\\
\hline
-2 & 52147.0    & ~4.5        & ~-2.9\\
 0 & 52547.5    & ~2.0        & ~~0.0\\
 1 & 52748.3    & 11.4        & ~~2.0\\
 2 & 52943.5    & ~4.0        & ~-1.6\\
 3 & 53134.8    & ~4.0        & ~-9.1\\
 5 & 53529.7    & ~6.0        & -11.8\\
 7 & 53941.2    & ~3.5        & ~~2.1\\
 8$^{\star}$ & 54177.8    & 30.0        & ~39.9\\
11 & 54730.0    & ~9.5        & ~-4.3\\
12 & 54939.9    & 15.0        & ~~6.8\\
13 & 55123.0    & ~6.5        & ~-8.9\\
\hline
\end{tabular}
\begin{list}{}{}
\item[$^{\star}$] {Not used for timing analysis because of high uncertainty.}
\end{list}
\end{table}

The resulting pulsation period agrees very closely with the period
calculated using Fourier analysis.  The final ephemeris for pulsation maxima
(Eq.~\ref{pulslinefem}) can be used to postdict the moment of the local
maximum that indeed occurred during the last eclipse, with an accuracy of a
few days, corresponding to the pulsation epoch $E_{\mathrm p} = 10$ (see the
dotted arrow in Fig.~\ref{ASAS3.VI}).  It is worth noting that the
photographic observations of the eclipse at epoch $E\,=\,1$ show an
analogous, local pulsational maximum (see Fig.\,\ref{LClp}).

%
%______________________________________________________________

\subsection{Spectral Energy Distribution (SED)}

\begin{table*}
\caption{The photometry used to calculate the spectral energy distribution
(SED).  The observational magnitudes and fluxes were dereddened using
a~colour excess $E_{\rm{B-V}} = 0.46$.}
\label{sed.t}
\centering			% used for centering table
\begin{tabular}{lr|ll|ll|l}
\hline\hline
Band            & Wavelength           & \multicolumn{2}{c|}{Observational}    &\multicolumn{2}{c|}{Extinction corrected}   & Source\\
                & $\bar\lambda[$\AA$]$ & $[mag]$       & $\sigma$ & $[mag]$    & $Flux[erg$~$cm^{-2} s^{-1}]$ & \\
\hline
$B_T$           & 4220                 & 11.808        & 0.096    & 9.85       & 3.297e-09                    &TYCHO\\
$m_{\rm pg}$    & 4300                 & 11.4          & 0.2      & 9.48       & 4.555e-09                    &Swope 1936\\
$V_T$           & 5353                 & 10.656        & 0.062    & 9.20       & 4.279e-09                    &TYCHO\\
$V$             & 5450                 & 10.790        & 0.012    & 9.37       & 3.535e-09                    &ASAS-3\\
$I$             & 9000                 & ~9.667        & 0.034    & 9.04       & 1.727e-09		      &$-_{''}-$\\
$J$             & 12500                & ~8.631        & 0.020    & 8.29       & 1.752e-09                    &2MASS\\
$H$             & 16500                & ~8.072        & 0.027    & 7.87       & 1.279e-09                    &$-_{''}-$\\
$K$             & 22000                & ~7.613        & 0.016    & 7.49       & 8.644e-10                    &$-_{''}-$\\
\hline
                & $\bar\lambda[$\AA$]$ & $Flux[Jy]$    & $\sigma$ & $Flux[Jy]$ & $Flux[erg$~$cm^{-2} s^{-1}]$ & \\
\hline
9$\mu$m         & 90000                & 3.906e-01     & 0.082e-01& 3.942e-01  & 1.313e-10                    &AKARI PSC\\
12$\mu$m        & 120000               & 3.632e-01     & ...      & 3.656e-01  & 9.134e-11                    &IRAS FSC\\
12$\mu$m        & 120000               & 4.605e-01     & ...      & 4.636e-01  & 1.158e-10                    &IRAS PSC$^{\star}$\\
18$\mu$m        & 180000               & 2.511e-01     & 0.222e-01& 2.520e-01  & 4.197e-11                    &AKARI PSC\\
25$\mu$m        & 250000               & 2.796e-01     & ...      & 2.804e-01  & 3.363e-11                    &IRAS FSC\\
25$\mu$m        & 250000               & 1.118e+00     & ...      & 1.121e+00  & 1.344e-10                    &IRAS PSC$^{\star}$\\
60$\mu$m        & 600000               & 7.887e-01     & ...      & 7.893e-01  & 3.944e-11                    &IRAS FSC$^{\star}$\\
60$\mu$m        & 600000               & 5.477e-01     & ...      & 5.481e-01  & 2.739e-11                    &IRAS PSC$^{\star}$\\
100$\mu$m       & 1000000              & 5.090e+00     & ...      & 5.090e+00  & 1.526e-10                    &IRAS FSC$^{\star}$\\
100$\mu$m       & 1000000              & 1.182e+01     & ...      & 1.182e+01  & 3.544e-10                    &IRAS PSC$^{\star}$\\
\hline
\end{tabular}
\begin{list}{}{}
\item[$^{\star}$] {The IRAS PSC and FSC data have very low flux density
quality, and are best considered as upper estimates.}
\end{list}
\end{table*}

\begin{figure*}
  \centering
    \includegraphics[width=17cm]{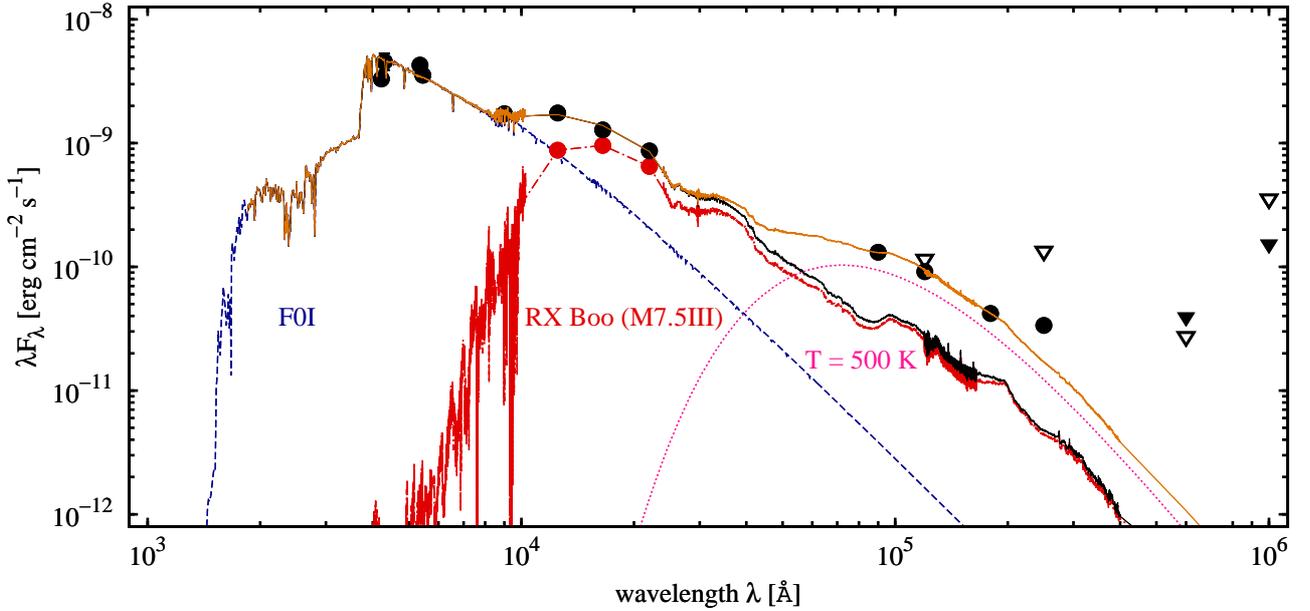}
\caption{Dereddened spectral energy distribution of V383\,Sco obtained from
the data given in Table \ref{sed.t} (points).  The dashed line (blue in the
electronic version of this paper) shows the F0I-type supergiant from the
1993 Kurucz Stellar Atmospheres Atlas \citep{Kur1993}.  The dot-dashed line
(red in the electronic version) marks the SED of the SRb-type pulsating star
RX\,Boo, built using the \emph{VOSpec} virtual observatory tool.  The dotted
line (magenta in the electronic version) shows a~black body with $T$=500\,K. 
The continuous line (orange in the electronic version) shows the sum of all
3 components appropriately rescaled (by a factor of $1.9 \times 10^{-20}$
for the Kurucz model of an F0I star and by a factor of $1.3 \times 10^{-4}$
for the observational SED of RX~Boo) to fit the SED of V383\,Sco.  Triangles
denote IRAS data with very low flux density quality (marked with the number
1 in the revised version of the IRAS Point Source Catalog), i.e.  upper
estimates: from the IRAS FSC catalogue (filled triangles) and from the IRAS
PSC catalogue (open triangles).}
  \label{SED.V383Sco.SCar.RXBoo}
\end{figure*}

There are not many existing photometric multicolour measurements of
V383\,Sco that can be used to construct its SED; however, these rare data
exist over quite a~wide spectral range.  The TYCHO-2 catalogue has $B_{\rm
T}$ and $V_{\rm T}$ magnitudes, although these are not very good quality. 
Inaccurate out-of-eclipse $m_{\rm pg}$ data is available from
\citet{Swo1936}.  The ASAS-3 survey gives values of $V$ and $I$ magnitudes
in the Johnson photometric system and 2-MASS $JHK$ observations complement
the energy distribution in the near-infrared.  The revised versions of the
Infrared Astronomical Satellite Faint Source Catalogue (IRAS FSC) and the
Point Source Catalogue (IRAS PSC) include a~potential counterpart for
V383\,Sco.  These data together with new AKARI satellite detections
significantly extend the observations to the mid- and far-infrared.  All the
available photometric out-of-eclipse flux estimations are given in Table
\ref{sed.t}.  The photometry was dereddened using the colour excess value
$E_{\rm{B-V}}=0.46$ (see section \ref{vel}) and by adopting the mean
interstellar extinction curve for $R=3.1$ developed by \citet{Fit2004}.  The
magnitudes from $B_{\rm T}m_{\rm pg}V_{\rm T}VIJHK$ bands were transformed
to the $\lambda F_{\lambda}$ fluxes using the \citet{Bes1998} calibration. 
The resulting fluxes are shown in Table~\ref{sed.t} and the SED is shown in
Figure~\ref{SED.V383Sco.SCar.RXBoo}.  The spectrum is dominated by an F-type
supergiant in the visual and there is a significant excess from the near- to
far-infrared.

\begin{table}
\caption{Luminosities of components of the SED of V383\,Sco.}
\label{LUMfromSED}
\centering			% used for centering table
\begin{tabular}{|l|rr|}
\hline\hline
% &\multicolumn{2}{c|}{SED with RX\,Boo} & \multicolumn{2}{c|}{SED with S\,Car}   \\
component of spectrum	& Luminosity $[erg s^{-1}]$ & $L L_{\sun}^{-1}$ \\%& Luminosity $[erg s^{-1}]$ & $L L_{\sun}^{-1}$ 
\hline
$L_{\mathrm {hot}}$     & 3.67 10$^{37}$ & 9566 \\ % & 3.67 10$^{37}$ & 9566
$L_{\mathrm {cool}}$    & 9.01 10$^{36}$ & 2346 \\ % & 8.13 10$^{36}$ & 2117
black body (500 K)      & 1.28 10$^{36}$ &  334 \\ % & 1.52 10$^{36}$ &  397
remnant at far IR       & $\sim$(0.8 -- 1.9) 10$^{36}$ & $\sim$(200 -- 500) \\ %& $\sim$(0.8 -- 1.9) 10$^{36}$ & $\sim$(200 -- 500)
\hline
\end{tabular}
%\begin{list}{}{}
%\item[$^{\star}$] ......................\\.................
%\end{list}
\end{table}

%
%______________________________________________________________

%%%%%%%%%%%%%%%%%%%%%%%%%%%%%%%%%%%%%
%%%%%%%%%%%%%%%%%%%%%%%%%%%%%%%%%%%%%

\subsection{Low-resolution spectrum} \label{secLRS}

\begin{figure}
   \resizebox{\hsize}{!}{\includegraphics{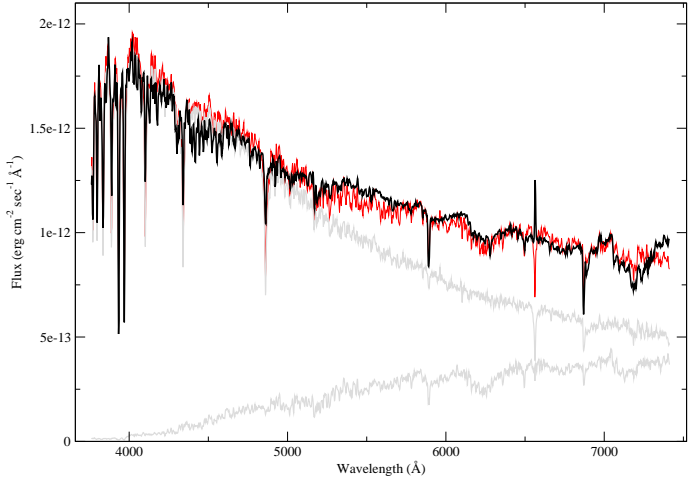}}
\caption{Observed low-resolution spectrum of V383\,Sco (thick line) compared
with the model (thin line) that is a~superposition of F0 and M1 supergiant
spectra (grey lines).  The observed spectrum and the model are dereddened by
$E_{\mathrm {B-V}} = 0.5$.}
  \label{V383Sco_low}
\end{figure}

A~first inspection of the low-resolution spectrum (LRS) of V383\,Sco
revealed that it is dominated by a~late A- or early F-type absorption
spectrum.  Only H$\alpha$ appears as a weak emission.  In the red part of
the spectrum, above $\lambda \sim 5500$\,{\AA}, traces of molecular bands
seem to be present.  It was obvious that the V383\,Sco spectrum is
a~superposition of a~relatively hot A--F spectrum and a~cool M spectrum.  In
an attempt to better define both components' spectra, we used the FITSPEC
task, from the STSDAS\footnote{STSDAS is a product of the Space Telescope
Science Institute, which is operated by AURA for
NASA.}.SYNPHOT\footnote{See~
\url{http://www.stsci.edu/institute/software_hardware/stsdas/synphot/SynphotManual.pdf}}
package in IRAF, with three free variables --- the renormalization of the
two spectra and $E_{\mathrm {B-V}}$.  The task searches for values of those
variables that minimize the residuals between the templates and the observed
spectrum.  As templates we used the spectra from the Jacoby-Hunter-Christian
spectrophotometric atlas \citep{Jac1984} which have a~resolution similar to
that of our spectrum.  The best model was obtained for a~superposition of
F0\,I and M1\,I spectra with a flux ratio $F_\mathrm{F} / F_\mathrm{M} =
3.9$ reddened by $E_{\mathrm {B-V}} = 0.5$.  The observed spectrum of
V383\,Sco and the fitted model are compared in Fig.~\ref{V383Sco_low}.  The
differences between the models for spectra in the ranges A9\,I -- F3\,I and
M1\,I -- M2\,I (not every spectral subclass has a template in the atlas) are
very small and the estimate of $E_{\mathrm {B-V}}$ varies from 0.4 to 0.6. 
Based on the LRS, we can infer that the spectra of the components visible in
the V383\,Sco spectrum are F0\,I and M1\,I with an accuracy of the order of
one or two subclasses and that the reddening in the direction of the star is
$E_{\mathrm {B-V}} = 0.5 \pm 0.1$.

%
%______________________________________________________________

\subsection{High-resolution spectra}

\begin{figure}
  \resizebox{\hsize}{!}{\includegraphics{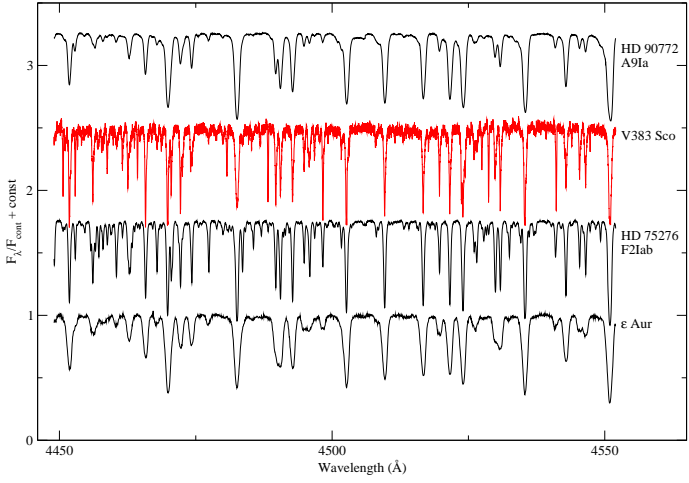}}
  \caption{HARPS spectrum of V383\,Sco in the region 4450--4550\,{\AA}
  compared with the spectra of HD\,90772 and HD\,75276 from the UVES library
  \citep{Bag2003} and $\varepsilon$\,Aur from the ELODIE archive
  \citep{Moultaka2004}.}
  \label{V383Sco_comp}
\end{figure}

A~careful examination of the HARPS spectra of V383\,Sco obtained on 2009
October 14 and 16 did not show any noticeable night-to-night changes.  Thus,
we average the high-resolution spectra, obtaining a high-resolution spectrum
(hereafter, HRS) with a higher S/N ratio.  Comparing the V383\,Sco spectrum
with those from the UVES library of high-resolution spectra \citep{Bag2003}
we found that it falls somewhere between the A9\,I and F2\,I spectral
classes (Fig.~\ref{V383Sco_comp}).  The spectrum of V383\,Sco is very close
to one of the spectra of $\varepsilon$~Aur from the ELODIE database
\citep{Moultaka2004}, obtained outside the eclipse on 2003 November 1
(Fig.~\ref{V383Sco_comp}).  Recently, \citet{Hoa2010} classified the
spectrum of the F star in $\varepsilon$~Aur as F0\,II--III?  (post-AGB), and
pointed out that it has the appearance of an F0 supergiant.  Based on the
HRS, we suggest an F0\,I spectral class for V383\,Sco as well.

\begin{figure}
    \resizebox{\hsize}{!}{\includegraphics{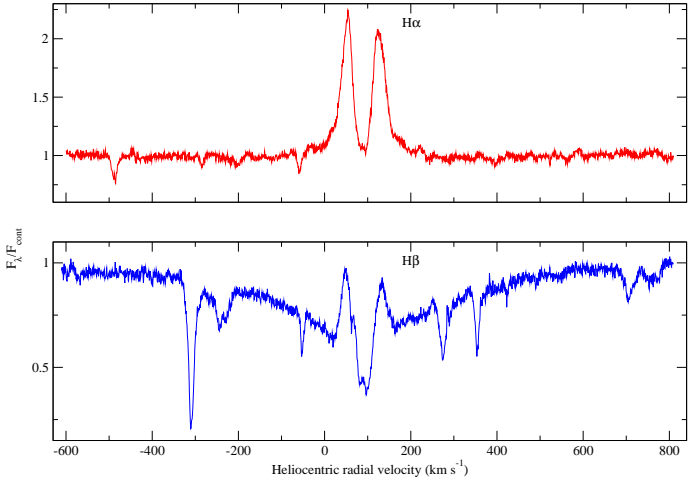}}
\caption{H$\alpha$ and H$\beta$ profiles in the HARPS spectrum of V383\,Sco.}
  \label{V383Sco_HaHb}
\end{figure}

The Balmer series lines appear similar to those in the spectrum of a~Be
star.  H$\alpha$ is present as two emission components divided by an
absorption (Fig.~\ref{V383Sco_HaHb}).  The emission components are separated
by about 70\,km\,s$^{-1}$ and the blue component is slightly stronger than
the red one.  The absorption seems to be a~blend of two components separated
by about 10\,km\,s$^{-1}$.  In H$\beta$, in addition to the same emission
and absorption components, the wings of a~wide absorption profile are
clearly visible (Fig.~\ref{V383Sco_HaHb}).  The radial velocities measured
on the H$\alpha$ and H$\beta$ emission wings are 89\,km\,s$^{-1}$ and
92\,km\,s$^{-1}$, respectively.  They are practically identical with the
velocity of the metallic absorptions (see Section~\ref{vel}).  In the
profiles of the higher series members, the wide absorption dominates. 
Traces of emission components in the core of the wide absorption are visible
only in H$\gamma$ and H$\delta$.

\begin{figure}
  \resizebox{\hsize}{!}{\includegraphics{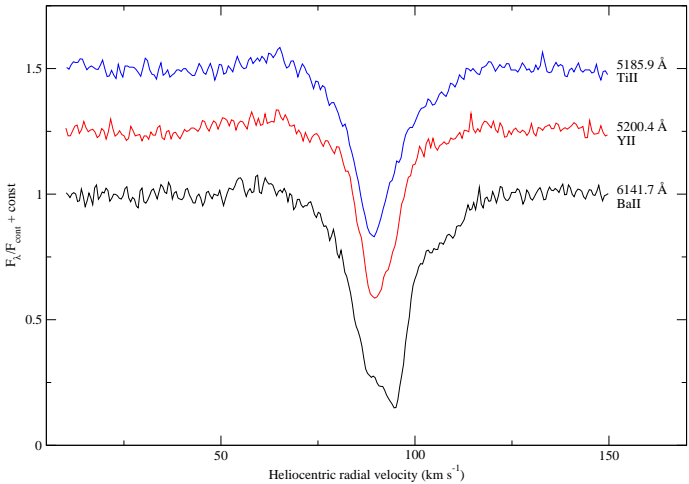}}
  \caption{Profiles of \ion{Ti}{ii} 5286\,\AA, \ion{Y}{ii} 5200\,\AA, and
\ion{Ba}{ii} 6142\,{\AA} in the spectrum of V383\,Sco in which weak
emissions on the shortward edges and wide absorption wings are clearly
visible.}
  \label{V383Sco_pcyg}
\end{figure}

\begin{figure}
  \resizebox{\hsize}{!}{\includegraphics{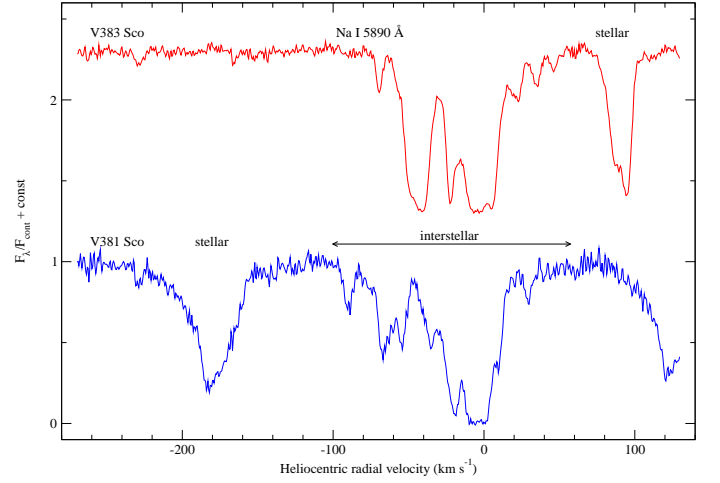}}
  \caption{Comparison of the region around \ion{Na}{i} 5890\,\AA\ in the
HARPS spectra of V383\,Sco (stellar absorption at around +90\,km\,s$^{-1}$)
and V381\,Sco (stellar absorption at about -180\,km\,s$^{-1}$).}
  \label{V383Sco_na}
\end{figure}

The most common features in the spectrum of V383\,Sco are sharp absorption
lines of neutral and singly-ionised metals.  Among the most conspicuous are
the \ion{Ba}{ii} absorptions.  All the lines of multiplets 1 and 2 in our
spectral region (4554\,\AA, 4934\,\AA, 5854\,\AA, 6142\,\AA, 6497\,\AA) are
present and are among the strongest features.  The \ion{Sr}{ii} lines
4078\,\AA\ and 4216\,\AA\ are strong as well.  In contrast, we were not able
to identify the \ion{Li}{i} line 6708\,\AA\ in our spectrum with S/N$\sim$35
in this region.  The strongest metallic absorptions show asymmetric or split
cores which indicate that they could be blends of two components (see for
example the \ion{Ba}{ii} 6142\,\AA\ line in Fig.~\ref{V383Sco_pcyg} and the
stellar \ion{Na}{i} absorptions in Fig.~\ref{V383Sco_na}).  Some lines show
weak, inverse P\,Cyg profiles with a faint blue emission component and wide
absorption wings (Fig.~\ref{V383Sco_pcyg}).  Such variety of structures
shows that the spectrum of V383\,Sco is very peculiar and more observational
material is necessary for its better understanding.

To estimate the projected rotational velocity of the F star we used the
method described by \citet{Car2011}.  We divided the spectrum of V383~Sco in
the region 4500--5500\,{\AA} into eight intervals, avoiding H$\beta$ and the
interchip gap around 5320\,\AA.  In each interval, the spectrum of V383~Sco
was cross-correlated with a synthetic spectrum ($T_{\mathrm{eff}} =
7500$\,K, $\log g = 1.5$, microturbulent velocity 1\,km\,s$^{-1}$) from the
POLLUX database \citep{Pal2010}.  To correct for the instrumental width,
telluric lines were used.  Because the macroturbulent velocity for
luminosity class I, estimated from Fig.  17.10 in \citet{Gra2005}, appeared
to be higher than the total broadening, we used the formula for luminosity
class II developed by \citet{HeMe2007}.  We found a projected rotational
velocity $v\,\mathrm{sin}i = 7.6 \pm 0.5$\,km\,s$^{-1}$ for the F star in
V383\,Sco.  The reliability of this estimation is qualitatively supported by
Fig.~\ref{V383Sco_comp}.  It is obvious in the figure that the lines in the
spectrum of V383\,Sco are sharper than those in the spectrum of HD\,75276. 
\citet{DeM2002} estimate $v\,\mathrm{sin}i = 9.8$\,km\,s$^\mathrm{-1}$ for
HD\,75276.

\begin{figure}
  \resizebox{\hsize}{!}{\includegraphics{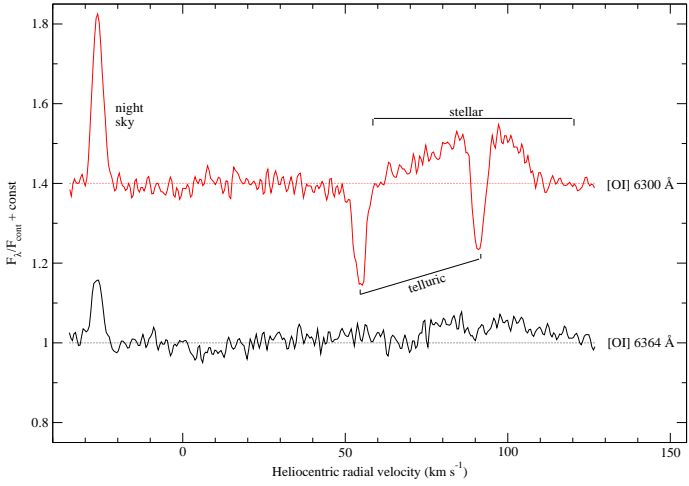}}
  \caption{[\ion{O}{i}] 6300\,\AA\ and 6364\,\AA\ forbidden emissions in the
spectrum of V383\,Sco.}
  \label{V383Sco_OI}
\end{figure}

In addition to the emissions in the strongest Balmer lines and the very weak
emission components in the inverse P\,Cyg profiles in the HARPS spectrum of
V383\,Sco, we also identified the [\ion{O}{i}] 6300\,\AA\ and 6364\,\AA\
forbidden emissions (Fig.~\ref{V383Sco_OI}).  The 6300\,\AA\ line is blended
with two telluric lines, but is clearly visible while the 6364\,\AA\ line is
very weak.  Both [\ion{O}{i}] forbidden emissions are shifted longward with
a velocity of $\sim 90$\,km\,s$^{-1}$ which equals the radial velocity of
the absorption lines in the spectrum of V383\,Sco (see Section~\ref{vel}). 
A~Gaussian fit to the [\ion{O}{i}] 6300\,\AA\ emission gives a~FWHM of the
order of $31$\,km\,s$^{-1}$.

%
%______________________________________________________________

\subsection{Reddening and distance to V383\,Sco and V381\,Sco --- two
objects located near the Galactic centre} \label{vel}

Like V383\,Sco ($l = 352\fdg9723,\ b = 06\fdg0999$), V381\,Sco ($V =
12\fm3,\ l = 354\fdg2974,\ b = 03\fdg8119$) is a poorly-studied
system.  There are wide discrepancies in the spectral classification of this
star in the literature.  Its discoverer, Henrietta Swope, assigned it
a~spectral type F0 and an unknown luminosity class \citep{Swo1936}, while
\citet{Pop1948} classified it as an A5\,Ia star.  In \citet{Bowers1978},
V381\,Sco is listed as an M5\,Ia red supergiant.  Comparing the spectrum of
V381\,Sco with the spectra from the UVES library of high-resolution spectra
\citep{Bag2003} we found the best match with an A8\,II spectrum.

In Fig.~\ref{V383Sco_na} we compare the region around the \ion{Na}{i}
D$_{\mathrm 2}$ 5890\,\AA\ line in the spectra of the two stars.  It is
evident that despite the proximity of the stars in the sky, the \ion{Na}{i}
interstellar line absorptions are notably different.  In both spectra there
are some components with similar radial velocities in the range from $-30$
to $20$\,km\,s$^{-1}$, which most probably originate in the same
interstellar clouds.  Several absorptions with different radial velocities
visible in the spectra indicate that additionally, there are different
interstellar clouds in the direction to each star.  By fitting Gaussians to
the particular absorption components, we measured the total equivalent width
of the interstellar \ion{Na}{i} D$_{\mathrm 2}$ absorptions.  Using the
calibration of \citet{Mun1997}, we estimated a~reddening $E_{\mathrm {B -
V}} = 0.43$ for V383\,Sco which agrees with the value obtained from the
fitting of the LRS.  For V381\,Sco we obtained a~reddening value $E_{\mathrm
{B - V}} = 0.39$.  According to \citet{Mun1997}, the accuracy of the
reddening estimation based on the \ion{Na}{i} equivalent width is higher for
$E_{\mathrm {B - V}} \leq 0.4$.  For multi-component profiles of
\ion{Na}{i}, as in our case, the accuracy is generally $\sim 0\fm15$.  The
resulting estimate of $E_{\mathrm {B - V}}$ is only a lower limit, since the
\ion{Na}{i} lines can be saturated.  Hereafter, we use an average
$E_{\mathrm {B - V}} = 0.46 \pm 0.15$ for V383\,Sco.

Assuming an average out-of-eclipse brightness of V383\,Sco $V = 10\fm8$,
a~reddening $E_{\mathrm {B - V}} = 0.46 \pm 0.15$, and using $M_{\mathrm V}
= 5\fm1$ for an F0\,Ib supergiant from \citet{SK1982}, we derive a~distance
to the star of $7.8^\mathrm{+1.9}_\mathrm{-1.5}$\,kpc.  If we use the
absolute magnitudes for F0\,Ia or Iab supergiants \citep{SK1982}, then the
estimated distance is unrealistic and exceeds 15\,kpc.  Using $E_{\mathrm {B
-V}} = 0.39 \pm 0.15$ (estimated above), an out-of-eclipse magnitude $V =
12\fm3$, and $M_{\mathrm V} = -2\fm6$ for an A8\,II bright giant \citep
{SK1982}, we estimate the distance to V381\,Sco to be
$5.5^{+1.3}_{-1.1}$\,kpc.

When comparing the spectra of V383\,Sco and V381\,Sco we see an impressive
difference in the radial velocities of the stellar absorption lines.  This
difference can be easily seen in the stellar \ion{Na}{i} absorptions in
Fig.~\ref{V383Sco_na}.  Using about 350 metallic absorption lines in the
spectrum of V383\,Sco, we measured an average heliocentric radial velocity
$V_{\mathrm h} = 89.8 \pm 0.1$\,km\,s$^{-1}$.  Measuring about 190
absorption lines in the spectrum of V381\,Sco, we obtained an average
heliocentric velocity $V_{\mathrm h} = - 178.8 \pm 0.2$\,km\,s$^{-1}$.  The
corresponding velocities relative to the local standard of rest (LSR) are
$V_{\mathrm {lsr}} = 98$\,km\,s$^{-1}$ and $V_{\mathrm {lsr}} =
-171$\,km\,s$^{-1}$ for V383\,Sco and V381\,Sco, respectively.

The absolute velocities of both stars are high.  They cannot be caused by
orbital motion in such long-period systems.  There is no evidence for their
having a high proper motion in the literature.  Most probably, these high
velocities are connected with the kinematics of the regions close to the
Galactic centre.  V383\,Sco and V381\,Sco lie close to the Galactic centre
and to the Galactic plane ($z \sim 0.8$\,kpc and $z \sim 0.4$\,kpc
respectively).  An examination of the Galactic-longitude--velocity diagrams
of CO emission \citep{Dame2001}, \ion{H}{i} \citep[see Fig.~1
in][]{Weiner1999}, Galactic planetary nebulae \citep{Durand1998} and OH/IR
and SiO-maser stars \citep{Hab2006} shows that the observed velocities of
V383\,Sco and V381\,Sco fall into regions close to the Galactic centre with
very high velocities.  The kinematics of the gas in these regions is
inconsistent with purely circular motion \citep[and references
therein]{Dame2001, Weiner1999}.  Thus, we cannot use the standard Galactic
rotation curve \citep[e.g.][]{Sof2009} to estimate the kinematic distances
of V383\,Sco and V381\,Sco.

Taking into consideration the distances to V383\,Sco and V381\,Sco estimated
above, as well as their Galactic coordinates and the measured velocities
relative to the LSR, both stars seem to belong to the bulge/bar structure in
the inner part of the Milky Way \citep[and references therein]{Weiner1999,
Hab2006, Vanhol2009}.  Moreover, the velocimetric model proposed by
\cite{Vall2008} suggests that the spatial kinematic characteristics of these
stars are inconsistent with their location in any of the inner spiral arms
of the Galaxy.

One way to roughly estimate the distance to V383\,Sco and V381\,Sco is to
use the kinematic model developed by \citet{Weiner1999} for the inner part
of our Galaxy.  From the velocity contour plots in Fig.~8 in their paper we
infer that V383\,Sco is placed about 0.5\,kpc behind the Galactic centre and
that V381\,Sco is placed about 1\,kpc in front of it.  For a distance to the
Galactic centre $R_{\mathrm 0} = 8.0$\,kpc \citep{Sof2009} the estimated
distances to V383\,Sco and V381\,Sco are 8.5\,kpc and 7\,kpc, respectively. 
The main sources of errors in these estimates are the determination of the
stars' positions in Fig.~8 of \citet{Weiner1999} by sight and the
uncertainty in the $R_{0}$ estimate \citep[see Table~1 in][]{Vanhol2009}. 
Thus, we estimate the combined random and systematic error of the kinematic
distances as $\pm 1$\,kpc.  Thus, these distances are statistically
consistent with the luminosity-based estimates indicated above.

To obtain the distance to V383\,Sco by a third, independent method, we used
the period--luminosity relation for pulsating Mira-type stars according to a
formula given by \citet{White2000}: $M_{\mathrm K} = -3.47 \log{P} + \beta$
($\beta = 0.84 \pm 0.14$).  Using $P_{\mathrm {pul}} = 198\fd8$ and
dereddened infrared brightness $K_0=7\fm49$ we obtained a distance modulus
$(K_0 - M_{\mathrm K}) = 14.63 \pm 0.20$ and a distance of $8.4 \pm
0.8$\,kpc.  This value agrees closely with the previous two estimates
($7.8^\mathrm{+1.9}_\mathrm{-1.5}$\,kpc from the $M_{\mathrm V}$ calibration
and $8.5 \pm 1$\,kpc from the kinematic model of the Galactic centre). 
Considering these three estimates to be independent Gaussian distributions
and taking a weighted mean (conservatively using 1.9\,kpc as the standard
deviation for the $M_{\mathrm V}$ estimate), we obtain $8.4 \pm 0.6$\,kpc as
the distance to V383\,Sco.  Similarly, the weighted mean distance to
V381\,Sco from the reddening (conservatively using $5.5\pm1.3$\,kpc) and
kinematic ($7\pm1$\,kpc) distances is $6.4\pm0.8$\,kpc.

%%%%%%%%%%%%%%%%%%%%%%%%%%%%%%%%%%%%%%%%%
%%%%%%%%%%%%%%%%%%%%%%%%%%%%%%%%%%%%%%%%%

%
%______________________________________________________________

\subsection{Discussion of the model of V383\,Sco} \label{simplemodel}

Our interest in V383\,Sco arose from the suspicion that it could be a~system
similar to EE\,Cep and $\varepsilon$\,Aur --- unique long-period eclipsing
binaries with a~dusty debris disk as a component causing eclipses
\citep{MiGr1999}.  We considered the possible similarity between V383\,Sco
and two competing high- and low-mass models of $\varepsilon$\,Aur \citep[see
e.g.][]{Gui2002}.  The analysis of the SED by \citet{Hoa2010} and
conclusions drawn from interferometric observations of the disk movement
relative to the F-type star by \citet{Klo2010} seem to validate the low-mass
model of $\varepsilon$\,Aur with a~single B5V-type primary embedded in a
dusty disk and a~much less massive F-type post-AGB secondary.  However,
there still exist strong arguments in favour of the model with a high mass
F-type supergiant and perhaps a binary system at the disk centre
\citep[see][]{Cha2011}.  We found that only the low-mass variant can be
applicable to V383\,Sco which explains the long-lasting eclipses ($\sim
11\%$ of the orbital period) that are observed.  Nevertheless, one
significant observational constraint strongly disagrees with the low-mass
$\varepsilon$\,Aur model.  The depth of the V383\,Sco eclipses shows
a~strong dependence on the mean wavelength of the photometric band, which is
clearly visible in the $V-I$ colour index (Fig.~\ref{ASAS3.ecl.VI}).  The
colour changes during the eclipses exceed 2~\fm0, where the eclipse in the
$I$ band is more than twice as shallow as it is in the $V$ band, which
suggests that the eclipsing body has to contribute significantly to the
total flux in the near-infrared.  This case is not similar to the systems
with a dark, eclipsing dusty debris disk like $\varepsilon$\,Aur and
EE\,Cep.  Instead, the eclipsing object has to be a cool supergiant.  A good
alternative that satisfies this condition is the resemblance to the BL\,Tel
system.  In BL\,Tel the eclipses caused by an M-type supergiant with orbital
period $P_{\mathrm orb}=$778~\fd0 show a similar asymmetric shape to those
of V383\,Sco and have a similar duration of about 10$\%$ of the orbital
phase \citep[see][]{vGen1986}.  Pulsations with a period of 65\fd1 observed
in BL\,Tel correspond to a hot component, a UU\,Her-type variable
\citep{vGen1977} --- a~subtype of SRd-type variables.  In the case of
V383\,Sco, however, the amplitude of the 200-day variations is higher for
longer wavelengths, which has not been observed in any type of pulsating
stars and is unclear given the brightness changes caused by pulsations. 
Thus we conclude that the pulsating component of the V383\,Sco system cannot
be a hot star, but rather a~cool component dominating in the IR.

Another, more likely possibility is that V383\,Sco is similar to HD\,172481
which, according to \citet{Reyn2001}, consists of an F-type post-AGB star
and a cool M-type companion, probably AGB.  \citet{White2001} showed that
the cool component in this system is a~Mira-type variable with a~pulsation
period of 312$^\mathrm{d}$.

Several arguments can be considered in favour of the post-AGB nature of the
F0\,I component in V383\,Sco: (i) the system's location in the bulge/bar
structure of our Galaxy; (ii) the high radial velocity; and (iii) the
presence of dust that would explain the IR excess in the SED
(Fig.~\ref{SED.V383Sco.SCar.RXBoo}).  Additionally, the [\ion{O}{i}]
6300\,\AA\ emission in the spectrum of V383\,Sco may indicate the presence
of a very low excitation nebula.  If we assume that the wide and asymmetric
profile of this forbidden line originates in an expanding nebula, then the
observed outflow velocity will be of the order of 15\,km\,s$^{-1}$, which is
typical for post-AGB expanding envelopes.  Nevertheless, the infrared excess
observed in the SED is a~little too low to be a~post-AGB star.

However, much stronger arguments favour the hypothesis that there is a~cool,
pulsating M-type component in the V383\,Sco system.  Two arguments follow
from the spectral energy distributions, on the basis of spectra and/or
photometric data: (i) In Fig.~\ref{V383Sco_low} in the red part of the
low-resolution spectrum there are traces of molecular absorption band
features characteristic of an M-type supergiant.  The spectrum in the visual
domain can be fitted with a~combination of an F0I and an M1I spectrum.  (ii)
There is an infrared excess visible in the SED
(see~Fig.~\ref{SED.V383Sco.SCar.RXBoo}).  At first glance, one might suspect
that the system includes a very cool non-stellar component as the companion
of the F-type star that cause the eclipses.  If we apply the black-body
approximation to the SED in the near IR part, it may seem that the observed
excess could not be produced by a cool star.  However, as a~consequence of
efficient mass loss accompanied by dust creation, pulsating Miras or
semiregular SR-type stars have SEDs with maxima strongly shifted to the near
infrared with respect to a black body of the same temperature
\citep[see][]{Lob1999}.  Thus the observed SED of V383\,Sco can be
reproduced by a superposition of two stellar spectra --- a hot supergiant of
F0I-type and a cool M-type giant/supergiant.  In
Figure~\ref{SED.V383Sco.SCar.RXBoo} we compared an~SED of V383\,Sco with the
sum of the two spectra, individually rescaled to match the observed absolute
flux levels, and a black body with a temperature of 500 K to explain the
excess in the range up to 18\,$\mu$m.  The F0I-type spectrum is from the
1993 Kurucz Stellar Atmospheres Atlas \citep{Kur1993}.  The spectrum of the
SRb-type star RX\,Boo (Fig.~\ref{SED.V383Sco.SCar.RXBoo}) was built using
the \emph{VOSpec} virtual observatory tool
\footnote{{\protect{\url{http://www.sciops.esa.int/index.php?project=SAT&page=vospec}}}}. 
According to the GCVS database RX\,Boo is an SRb-type star that varies with
an amplitude of $\sim$2~\fm5 in the $V$-band and changes spectral type at
least in the range M6.5e--M8IIIe with a typical period of $P_{\mathrm
{pul}}=162$~\fd3.  The presence of a cool star such as RX\,Boo is sufficient
to reproduce the observed SED of the V383\,Sco system.

There ara two more arguments for the presence of an M-type supergiant in the
V383\,Sco system, which provide additional evidence for pulsations of this
component, come from the photometric behaviour.  (iii) The eclipse depth
strongly depends on the colour, i.e.  on the mean wavelength of the
photometric band, which shows that a cool and very bright object dominates
in the near IR.  (iv) The observed amplitude of pulsations is much smaller
at shorter wavelengths.  Using the out-of-eclipse light curves, we estimate
amplitudes of the pulsation related variations as 0~\fm41 and 0~\fm21 for
the $I$ and $V$ bands, respectively (see Table~\ref{t.f}).  Normally the
amplitude of pulsations decreases towards longer wavelengths, but in this
case it is the opposite.  This can be explained by the amplitude of
pulsations in short wavelengths being substantially ``suppressed'' by the
presence of a much brighter (in the visible range) F-type supergiant.  We
are not sure how deep the eclipses really are because the minima were not
covered during the eclipses.  However, by extrapolating the changes near the
mid-eclipse during the most recent epoch we estimate that the apparent
magnitudes during the minimum should not be greater than about 11~\fm1 and
14~\fm3 for the $I$ and $V$ bands, respectively.  If we estimate the mean
magnitudes in the minima of pulsations as 9~\fm83 and 10~\fm86 for $I$ and
$V$, and assume that eclipses are total, we can estimate that the flux
ratios of both components $F_{\rm M}/F_{\rm F}$ change during the pulsation
cycle in the range 0.444--1.107 in $I$ and 0.044--0.267 in $V$.  The upper
value 0.267 agrees closely with the upper value from the LRS in the visible
domain obtained around the maximal phase of pulsations: $F_{\rm F}/F_{\rm M}
= 3.9 \Rightarrow F_{\rm M}/F_{\rm F} = 0.256$ (see Sec.~\ref{secLRS}). 
Hence, the amplitudes of pulsations of the cool M~component could reach
about $1$~\fm0 and $2$~\fm0 in the $I$ and $V$ bands, respectively, stronger
in the shorter wavelengths as is expected for a pulsating star.  In this
case the cool component could be a~semiregular, SR-type pulsating star.  The
estimated amplitude of pulsations strongly depends on the depths of
eclipses.  With increasing eclipse depth, the ratio $F_{\rm M}/F_{\rm F}$
will decrease and this is reflected by a~higher amplitude of pulsations of
the M-type component.  Indeed, the observed SED suggests that the ratio of
the components fluxes $F_{\rm M}/F_{\rm F}$ in the $I$ band could be a few
times (roughly an order of magnitude) lower than estimated from the
photometric data.  The true amplitude of pulsations could in fact be much
higher and the pulsating component could well turn out to be a~Mira-type
star.  Very high amplitude pulsations could provide an explanation for the
changes of eclipse contact moments observed during different cycles, which
can result from the variations of the pulsating supergiant radius.  It could
also explain the observed local maxima, visible during eclipses (e.g.  at $E
= 1$ and $E = 8$) as a consequence of an increase in the cold component's
flux (caused by pulsations) together with a decrease of the hot component's
flux (caused by obscuration).

\begin{table}
\caption{Estimated minimum values of the eclipse depths ($M_{\rm
midecl}^{\star} - \bar{M}_{\rm out}$) and the pulsation amplitudes
$\delta$.}
\label{t.f}
\centering			% used for centering table
\begin{tabular}{llllll}
\hline\hline
Band     & $\delta$ & $\delta_{\rm min}$ & $\delta_{\rm max}$  & $\bar{M}_{\rm out}$ & $M_{\rm midecl}^{\star}$\\
% & $Bump_{\rm max}$\\
\hline
$I$      & 0.41  & +0.15  &  -0.26 & ~9.68 & 11.1\\
% & 9.95\\
%$V-I$    & 0.23  & +0.11  &  -0.12 &       &      &\\
$V$      & 0.21  & +0.07  &  -0.14 & 10.79 & 14.3\\
%& 12.5\\
\hline
\end{tabular}
\begin{list}{}{}
\item[$^{\star}$] from extrapolation
\end{list}
\end{table}
%JDgarb_max = 2454536
%JDecl_min  = 2454489

\begin{figure}
  \resizebox{\hsize}{!}{\includegraphics{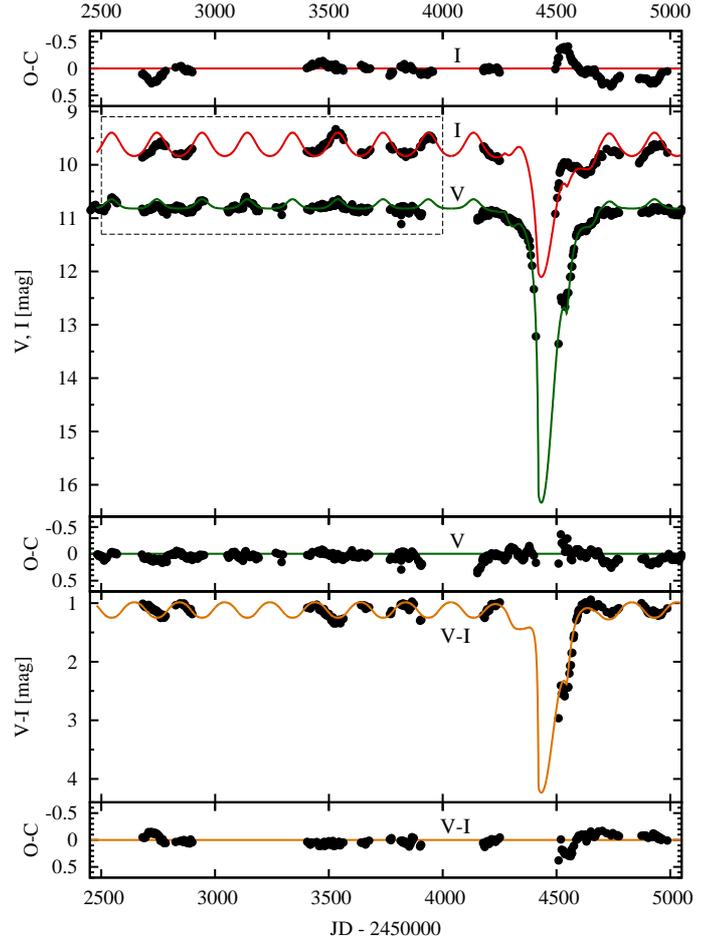}}
  \caption{Synthetic $V$- and $I$-band light curve and $V - I$ colour index
  curves obtained by modelling, plotted with solid lines (green, red, and
  orange, respectively in the electronic version of this paper), in
  comparison with observational photometric data.  Between the flux and
  colour curves, residuals are shown.  The model assumes a temperature of
  the cool star $T_{\mathrm {cool}}$=2800\,K.  The expanded view of the
  out-of-eclipse $V$ and $I$ light curves (box with a dashed line) is shown
  in Fig.~\protect\ref{expLCVI}.}
  \label{Model.ecl.VI}
\end{figure}

\begin{figure}
  \resizebox{\hsize}{!}{\includegraphics{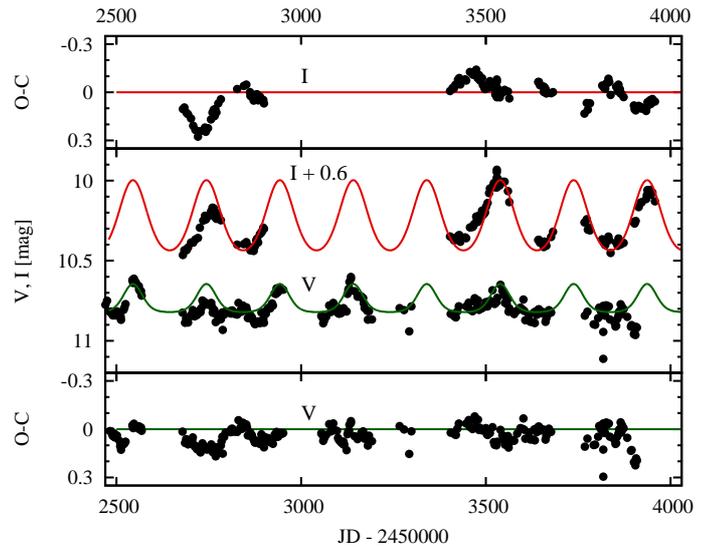}}
  \caption{Expanded view of the synthetic fit to the out-of-eclipse $V$ and
$I$ light curves and residuals.}
  \label{expLCVI}
\end{figure}

We have prepared a~model of the 2007/8 eclipse of V383\,Sco taking into
account the pulsations of the cool component using our own simple computer
code.  Our model is extremely simplified --- the stars are considered as
ideal spheres, limb darkening has been neglected, and stellar fluxes are
approximated as black bodies.  The changes in the radius of the M-type
supergiant and in its brightness have been expressed with cosine functions. 
A more detailed description of the model assumptions and the parameters that
have been derived are presented in the online Appendix \ref{AppendixM}.  As
inputs to the model we assumed that the hot component has an effective
temperature $T_{\mathrm {cool}} = 7500$\,K and its radius $R_{\mathrm {hot}}
= 58\,R_{\sun}$ was estimated using the Stefan-Boltzman law by adopting the
luminosity $L_{\mathrm {hot}}$ extracted from the SED
(see~Table~\ref{LUMfromSED}).  The best solution was found for the M
supergiant at effective temperature $T_{\mathrm {cool}} = 2800$\,K and with
radius $\bar{R}_{\mathrm {cool}} = 208\,R_{\sun}$, which is changing due to
the pulsations by $\Delta R$ = $\pm$0.19$\bar{R}_{\mathrm {cool}}$.  In our
model the supergiant totally obscures the hot component during an eclipse
minimum.  The resulting amplitudes of the brightness changes in the cold
supergiant --- if it were observed as a~solitary star --- would be 3\fm66
and 1\fm75 in the $V$ and $I$ bands, respectively.  The synthetic $V$- and
$I$-band light curves and $V-I$ colour index curve are shown in
Figure~\ref{Model.ecl.VI} and compared with photometric observational data. 
The eclipse is very deep, about 5\fm6 in $V$ and 2.5 in $I$ and needs to be
confirmed by future observations, because the system has never been observed
at the exact moment of minimum.  It will be necessary to organize an
observational campaign for the next eclipse in order to obtain data with a
dense sufficiently time coverage.  With the parameters of the system
components derived from our model, we estimate that if a secondary eclipse
occurs, it would be very shallow, i.e.  a few 0\fm001 in the $V$ band and
not deeper than about 0\fm02 in $I$.  Given the strong changes in brightness
caused by pulsations that are quite irregular in terms of amplitude it is
rather certain that the secondary eclipse will be undetectable in the
visible range.  The situation is somewhat better in the near-infrared ($JHK$
bands), where we estimate that the eclipse depths should fall in the range
0\fm04--0\fm06, though this would still be very difficult to observe. 
Careful inspection of Figure~\ref{expLCVI} reveals an additional interesting
feature in the synthetic light curves.  To our surprise and satisfaction,
our model has reproduced flattened pulsational minima in the synthetic
$V$-band light curve, which we had noticed in the phased observational
$V$-band light curve (see~Fig.~\ref{Phased.outofecl.VI.1p},~top) without
understanding the origin of this phenomenon.  Now we know that this is the
natural consequence of the fact that the pulsating star is a component of a
system with an object which strongly dominates in a~given photometric band. 
The changes which look roughly sinusoidal for a~single pulsating star will
have flattened minima when we put the pulsating star in a~binary system
together with a~much brighter star\footnote{Let us assume that a variable
pulsating star shows roughly sinusoidal variations in apparent magnitudes
$V_{\mathrm C}(t)\sim\sin{f(t)} = -2.5\log(F_{\mathrm C}(t))$.  When we
insert this star into the binary system with a non-variable companion, the
system as a whole will change its brightness according to the expression
$V_{\mathrm S}(t)\sim -2.5\log{(10^{-0.4\sin{f(t)}}+F_{\mathrm H})}$, where
$F_{\mathrm H}$ denotes the contribution of the companion to the total flux,
and the total flux of the system at any time is $F_{\mathrm S}(t) =
F_{\mathrm C}(t) + F_{\mathrm H}$.  When the contribution of the companion
to the total flux of the system is small $(F_{\mathrm H} \ll F_{\mathrm
C}(t) = 10^{-0.4\sin{f(t)}})$, the system shows almost sinusoidal
variations.  In contrast, when $(F_{\mathrm H} \gg F_{\mathrm C}(t)$), the
$V$ light curve consists of broad minima alternating with relatively short
maxima.}.  This result can constitute the additional, fifth (v) argument to
support the proposed model of V383\,Sco with a~pulsating cool supergiant as
the eclipsing component.  The best results are obtained with a~low
temperature of the cool component, $T_{\mathrm {cool}} = 2800$\,K, which
agrees very closely with the result for the M star based on the SED.  The
unreddened $V - I$ colour index close to the mid-eclipse is approximately
4\fm0, which is close to the index expected for a~M7-8 star.  We found
earlier (Sect.~\ref{secLRS}, Fig.~\ref{V383Sco_low}) that the M star was
better described by an M1-type fit to the low-resolution spectrum.  The
difference between M1 and M7-8 is significant and corresponds to a
temperature difference of about 700--800\,K.  On the other hand, the
low-resolution spectrum was obtained at a~pulsational maximum (on 31~Oct. 
2009 -- JD\,2455136), according to the ephemeris from
(Eq.~\ref{pulslinefem}) at a~pulsation phase almost exactly equal to zero
($\sim$0.03).  At that time, the star had reached the~temperature close to
the maximum possible value of the pulsation cycle.  Thus, the results are
consistent with our model, because during its pulsations, the cool star
changes its spectral type, at least in the range from about M1 to about
M7--8.

%
%______________________________________________________________

\section{Conclusions}

Using the ASAS-3 photometric data together with high (HARPS) and low
(1.9~m~telescope at SAAO) resolution spectra, we have revised our knowledge
of the long-period eclipsing binary V383\,Sco, for the first time since the
1930s.  Contrary to our initial expectations, the system does not resemble
the unique systems with dusty debris disks as eclipsing bodies, neither the
famous, very long period eclipsing binary $\varepsilon$\,Aur, nor EE\,Cep. 
Instead, we have found a~number of arguments in favour of a~new model in
which V383\,Sco could be similar to HD\,172481 --- a system with a post-AGB
F0-type star and a cool, pulsating M-type supergiant.  The most important
difference between these systems lies in the orientation of the orbits which
enables the observation of eclipses only in the case of V383\,Sco.  While
the hot component in V383\,Sco shows some features characteristic of
post-AGB stars, the detected infrared excess is too small and we can say
with certainty that it is a~supergiant of spectral type approximately F0. 
However, in the case of the second component that causes the eclipses, we
found strong evidence that it is a cool, pulsating M-type supergiant on the
basis of the following arguments:\\
i) Traces of molecular absorption bands, characteristic of an M-type
supergiant, are present in the red part of the low-resolution spectrum\\
ii) The observed SED can be reproduced by the superposition of two stellar
spectra, a~hot supergiant of F0I-type, and a~cool M-type giant/supergiant\\
iii) the eclipse depth strongly depends on the photometric band --- it
decreases in the direction of increasing wavelength\\
iv) The pulsation amplitude is lower in the shorter wavelength band\\
v) The flattened ``bottoms'' of the pulsation minima observed in the $V$
band are consistent with a synthetic model of the eclipses.\\
In addition, the presence of the pulsating supergiant in the system could
explain the changes in duration and shape of the eclipses.

The forbidden emission [\ion{O}{i}] 6300\,\AA\ in the HRS indicates the
presence of a~very low excitation nebula around the V383\,Sco system.\\

V383\,Sco and a~similar eclipsing binary V381\,Sco (by chance located close
to each other in the sky), have very high and oppositely directed radial
velocities: $89.8$ km\,s$^{\mathrm -1}$ vs $-178.8$ km\,s$^{\mathrm
-1}$.  These are in agreement with the kinematics of the regions close to
the Galactic centre.  V383\,Sco and V381\,Sco lie close to the Galactic
centre and to the Galactic plane ($z \sim 0.8$\,kpc and $z \sim 0.4$\,kpc,
respectively).  The distance to V383\,Sco estimated using reddening, radial
velocities, and the period-luminosity relation is 8.4\,$\pm$0.6\,kpc.  The
distance to V381\,Sco estimated from the reddening and kinematic distances
is $6.4\pm0.8$\,kpc.

The observational material collected so far is just sufficient to construct
a~rough model of the V383\,Sco system, which reproduces the main features of
observed photometric changes in the $V$ and $I$-band light curves.  With our
model we obtain very deep eclipses: about 5\fm6 in $V$ and 2\fm5 in $I$. 
This result requires confirmation by future observations, as the exact
moment of minimum has never been observed.

The next eclipse should start in the middle of 2020.  It is important to
carry out an extensive photometric and spectroscopic campaign to observe it. 
In the meantime, systematic spectroscopic monitoring would be useful to
obtain a spectroscopic orbit, with which it would became possible to obtain
a more precise estimate of the basic parameters of the system's components
such as their masses and radii.  Simultaneous photometric observations would
be still valuable for the study of changes in the M-type supergiant radius
as a function of the pulsational phase.  Observational material obtained for
this object over a decade time scale is still quite scant and any additional
data should lead to very valuable and interesting results.

%
%______________________________________________________________

\begin{acknowledgements}
This study was supported by MNiSW grants N203~018~32/2338 and N203~395534
and financial assistance was given to DG by the GEMINI-CONICYT Fund,
allocated to project 32080008.  WG and BP acknowledge financial support for
this work from the BASAL Centro de Astrofisica y Tecnologias Afines (CATA)
PFB-06/2007.  WG also gratefully acknowledges support from the Chilean
Center for Astrophysics FONDAP 15010003.  We greatly acknowledge the
variable star observations from the AAVSO International Database contributed
by observers worldwide, and used in this research.  This research was
conducted in part using the POLLUX database
(\url{http://pollux.graal.univ-montp2.fr}) operated at LUPM (Universit\'{e}
Montpellier II - CNRS, France) with the support of the PNPS and INSU.  MG
was financed by the GEMINI-CONICYT Fund, allocated to project 32110014.  The
authors thank the referee, Petr Harmanec, for his very constructive
comments.
\end{acknowledgements}

%
%______________________________________________________________

\bibliographystyle{aa} % style aa.bst

%
%________________________________________________________________
%\appendix
\Online
 
%
%________________________________________________________________
\begin{appendix}%First online appendix

\section{Simple model of V383\,Sco}\label{AppendixM}

\begin{figure}
  \resizebox{\hsize}{!}{\includegraphics{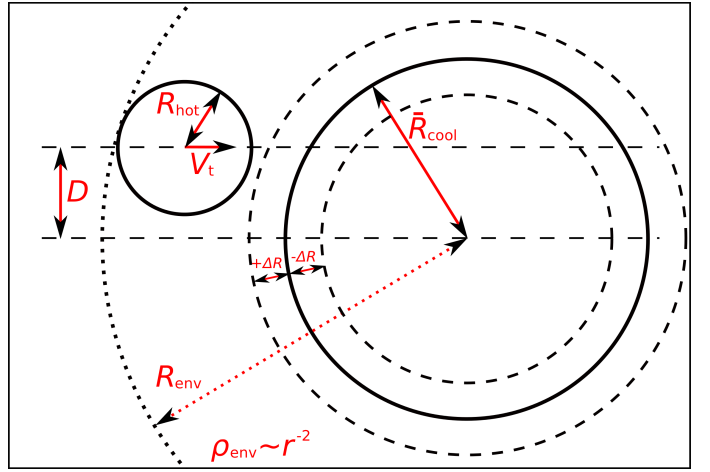}}
  \caption{Schematic explanation for the geometric parameters of the model.}
  \label{ModSchem}
\end{figure}

To make the model of the 2007/8 eclipse of V383\,Sco, which we describe
briefly below, many simplifying assumptions were made.  Stars were
considered as spheres with radii $R_{\mathrm {hot}}$, $\bar{R}_{\mathrm
{cool}}$ and effective temperatures of photosferes $T_{\mathrm {hot}}$,
$T_{\mathrm {cool}}$ for the hot F-type and cool M-type components,
respectively.  Limb darkening was neglected and stellar fluxes were
approximated as black bodies.  We describe the main eclipse (of the hot
component by the cool supergiant) by taking into account an impact parameter
$D$ which measures the projected distance between the centres of stellar
disks at the mid-eclipse point, as well as changes in the radius of the cool
star in the range $\pm \Delta R$ from the mean value of the radius
$\bar{R}_{\mathrm {cool}}$ as a~result of the pulsations
(Fig.\,\ref{ModSchem}).  Pulsational changes in the radius and brightness of
the cool component were described simply, similar to the approach used for
modelling pulsations in Mira $\chi$\,Cyg by \citet{ReGo2002}.  In our model
we assume that the changes in magnitudes and radius of the cool star can be
expressed with cosine functions

\begin{center} \begin{equation} M_{\mathrm {cool}} = \bar{M}_{\mathrm
{cool}} + \Delta M \cos{(\phi_{\mathrm {pul}})}, \label{cpM} \end{equation}
\end{center}

\begin{center} \begin{equation} R_{\mathrm {cool}} = \bar{R}_{\mathrm
{cool}} + \Delta R \cos{(\phi_{\mathrm {pul}} + \Delta \phi_{\mathrm
{pul}})}, \label{cpR} \end{equation} \end{center}

\noindent where $\Delta M$ is the semi-amplitude of brightness changes
caused by pulsations (from the mean value), $\phi_{\mathrm {pul}}$ is
a~pulsation phase, and $\Delta \phi_{\mathrm {pul}}$ is a~phase shift which
expresses by how much the moment of maximum radius of the M star precedes
the moment of maximum of its brightness.  The period and zero moment of
pulsation maxima were adopted from the ephemeris (Eq.\ref{pulslinefem}). 
Several parameters were treated as fixed, not subject to change in the
process of solution.  The effective temperature of the hot component
$T_{\mathrm {hot}}$ was set to 7500\,K and its radius $R_{\mathrm {hot}} =
58 R_{\sun}$ was estimated with the Stefan-Boltzmann law by adopting the
luminosity $L_{\mathrm {hot}}$ extracted from the SED
(see~Table~\ref{LUMfromSED}).  The broad, atmospheric parts of the eclipse
were included by introducing a light-absorbing envelope that changes the
density distribution as a function of distance from its centre $r$ as
$r^{-2}$.  Its radius $R_{\mathrm {env}}$ was roughly estimated from the
total duration of the atmospheric eclipse and fixed at 440\,$R_{\sun}$.  We
assumed that the intensity of radiation $\mathbb{I}$, measured after passing
through the envelope, changes from an initial value $\mathbb{I}_0$ according
to

\begin{center}
\begin{equation}  
\mathbb{I} = \mathbb{I}_0\,e^{-\tau (r)},  \label{II0}   
\end{equation}
\end{center}

where $\tau (r)$ is an~optical depth calculated

\begin{eqnarray}\label{tauu}
\tau(r)   &=& 2 \int_{\mathrm z=0}^{\sqrt{R_{\mathrm {env}}^2 - r^2} }\, \kappa\, \rho(r,z)\, dz \nonumber\\
          &=& 2\,C\,\kappa\,r^{-1} \arctan{\left(r^{-1} \, \sqrt{R_{\mathrm {env}}^2-r^2}\right)}.
\end{eqnarray}

\begin{figure}
  \resizebox{\hsize}{!}{\includegraphics{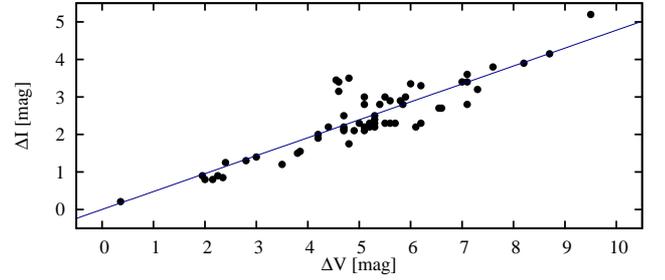}}
  \caption{Dependence of the pulsation amplitude of Mira/Semi-Regular
pulsating stars in the $V$ photometric band on the amplitude in the $I$ band
obtained from AAVSO light curves.}
  \label{delI_od_delV}
\end{figure}

The absorption coefficients $\kappa_{\mathrm V}$ and $\kappa_{\mathrm I}$
were chosen via visual comparison of synthetic curves with photometric
observational data and were fixed at the values 0.03 and 0.015 for $V$ and
$I$ bands, respectively, assuming that the integration constant $C$ is equal
to 1.  We carried out the solutions for several temperatures of the cool
component $T_{\mathrm {cool}}$ in the range from 2600\,K to 3500\,K with
a~step of 100\,K.  For each $T_{\mathrm {cool}}$ value we calculated the
star's radius $\bar{R}_{\mathrm {cool}}$ from the Stefan-Boltzmann law using
the luminosity extracted from the SED from the spectrum of RX\,Boo fitted in
the place of the cool component (see Table~\ref{LUMfromSED}).  The
adjustable, free parameters were: the moment of mid-eclipse defined as the
time at which the centres of stellar disks are at the minimum separation
$JD_{\mathrm 0}$, the reciprocal tangential velocity of the stars
$V_{\mathrm {t}}$, the semi-amplitude of brightness changes caused by
pulsations $\Delta M_{\mathrm V}$, $\Delta M_{\mathrm I}$ for $V$ and $I$
bands, respectively, and three parameters described earlier: $D$, $\Delta
R$, and $\Delta \phi_{\mathrm {pul}}$.  To reduce the number of free
parameters we combined the semi-amplitudes of $\Delta M_{\mathrm V}$,
$\Delta M_{\mathrm I}$ using an empirical relationship $\Delta M_{\mathrm I}
\approx 0.478\, \Delta M_{\mathrm V}$ that we found when analysing the AAVSO
photometric data for Miras- and SR-type stars (see Fig.~\ref{delI_od_delV}). 
The best solution, i.e.  corresponding to the minimum value of the sum of
square residuals, was obtained for $T_{\mathrm {cool}} = 2800$\,K and
$\bar{R}_{\mathrm {cool}} = 208\,R_{\sun}$.  The complete set of resulting
parameters of the model is shown in Table\,\ref{modpar}.

\begin{table}
\caption{Input (left) and output (right) parameters of the model of the
2007/8 eclipse of V383\,Sco.}
\label{modpar}
\centering			% used for centering table
\begin{tabular}{|lrl|lrl|}
\hline\hline
\multicolumn{3}{|c}{INPUT}					& \multicolumn{3}{|c|}{OUTPUT}\\
Parameter			& Value 	& Unit		& Parameter			& Value 	& Unit 		\\
\hline
$T_{\mathrm {hot}}$		& 7500		& K		& $T_{\mathrm {cool}}$		& 2800		& K		\\
$R_{\mathrm {hot}}$		& 58		& $R_{\sun}$	& $\bar{R}_{\mathrm {cool}}$	& 208		& $R_{\sun}$	\\
$R_{\mathrm {env}}$		& 440		& $R_{\sun}$	& $\Delta R$			& $\pm$0.19	& $\bar{R}_{\mathrm {cool}}$\\
$\kappa_{\mathrm V}$		& 0.03		& 1		& $D$				& 108		& $R_{\sun}$	\\
$\kappa_{\mathrm I}$		& 0.015		& 1		& $\Delta M_{\mathrm V}$	& 1.83		& mag		\\
				&		&		& $\Delta M_{\mathrm I}$	& 0.875		& mag		\\
				&		&		& $JD_{\mathrm 0}$		& 2454479	& day		\\
				&		&		& $V_{\mathrm {t}}$		& 1.349		& $R_{\sun}$\,day$^{-1}$ \\
				&		&		& $\Delta \phi_{\mathrm {pul}}$	& 0.24		& 1		\\
\hline
\end{tabular}
%\begin{list}{}{}
%\item[$^{\star}$] ...
%\end{list}
\end{table}

\end{appendix}

%
%________________________________________________________________

\begin{appendix} %Second online appendix

\section{Online photometric data}\label{AppendixT}

\addtocounter{table}{1}
\longtab{1}{
\small{
%\begin{landscape}
\begin{longtable}{lrlllrlllrll}
   \caption{\label{ASAS_V.dat} Photometric data obtained from the ASAS-3
survey in $V$ band.  The column labelled ``Grade" shows the quality flags,
where A denotes the highest quality of measurements and D the lowest.}\\
\hline\hline
$HJD-2450000$ & $V$ & Err$_V$ & Grade & $HJD-2450000$ & $V$ & Err$_V$ & Grade & $HJD-2450000$ & $V$ & Err$_V$ & Grade\\
\hline
\endfirsthead
\caption{continued.}\\
\hline\hline
$HJD-2450000$ & $V$ & Err$_V$ & Grade & $HJD-2450000$ & $V$ & Err$_V$ & Grade & $HJD-2450000$ & $V$ & Err$_V$ & Grade\\
\hline
\endhead
\hline
\endfoot
1947.88374 & 10.794 & 0.038 & A & 3059.85594 & 10.891 & 0.034 &  A & 4266.72275 & 10.926 & 0.048 & A\\
1950.86934 & 10.776 & 0.042 & B & 3062.88171 & 10.873 & 0.098 &  D & 4268.72660 & 10.965 & 0.047 & A\\
1954.87335 & 10.778 & 0.039 & A & 3067.89681 & 10.850 & 0.033 &  A & 4272.65519 & 11.012 & 0.030 & A\\
1955.85720 & 10.808 & 0.033 & A & 3075.86116 & 10.779 & 0.034 &  A & 4274.68990 & 11.022 & 0.031 & A\\
1961.88650 & 10.781 & 0.037 & A & 3080.85228 & 10.820 & 0.036 &  A & 4277.66524 & 11.004 & 0.029 & A\\
1963.86687 & 10.789 & 0.033 & A & 3086.89861 & 10.763 & 0.033 &  A & 4282.68026 & 10.999 & 0.038 & A\\
1965.86244 & 10.790 & 0.032 & A & 3090.83658 & 10.798 & 0.039 &  A & 4284.68381 & 11.024 & 0.032 & A\\
1967.86500 & 10.794 & 0.036 & A & 3096.91284 & 10.789 & 0.033 &  A & 4286.67763 & 11.072 & 0.040 & A\\
1978.81197 & 10.743 & 0.043 & A & 3099.83336 & 10.827 & 0.036 &  C & 4290.67641 & 10.996 & 0.031 & A\\
1979.84605 & 10.685 & 0.041 & A & 3102.85879 & 10.831 & 0.038 &  A & 4292.65072 & 10.997 & 0.030 & A\\
1980.82505 & 10.723 & 0.043 & A & 3107.89134 & 10.828 & 0.035 &  A & 4294.65876 & 11.038 & 0.038 & A\\
1981.84012 & 10.714 & 0.040 & A & 3113.80802 & 10.817 & 0.032 &  A & 4296.64837 & 11.055 & 0.034 & A\\
1982.80377 & 10.713 & 0.043 & A & 3114.71911 & 10.796 & 0.040 &  A & 4298.65374 & 11.031 & 0.032 & A\\
1994.78866 & 10.809 & 0.045 & B & 3116.79617 & 10.806 & 0.034 &  A & 4300.61757 & 11.015 & 0.042 & A\\
1997.80247 & 10.775 & 0.048 & B & 3122.82749 & 10.817 & 0.030 &  C & 4302.66254 & 11.059 & 0.030 & A\\
2025.75974 & 10.857 & 0.038 & A & 3125.75038 & 10.707 & 0.037 &  A & 4304.61034 & 11.053 & 0.029 & A\\
2032.76086 & 10.839 & 0.036 & A & 3127.74793 & 10.681 & 0.043 &  A & 4310.60157 & 11.067 & 0.033 & A\\
2033.75875 & 10.845 & 0.037 & A & 3129.79093 & 10.673 & 0.052 &  B & 4312.59860 & 11.049 & 0.038 & A\\
2080.59240 & 10.825 & 0.051 & D & 3130.88156 & 10.632 & 0.037 &  A & 4315.60113 & 11.077 & 0.029 & A\\
2083.64911 & 10.840 & 0.043 & A & 3134.80681 & 10.613 & 0.033 &  A & 4317.65819 & 11.066 & 0.030 & A\\
2087.63899 & 10.874 & 0.048 & B & 3134.81947 & 10.601 & 0.033 &  A & 4329.53502 & 11.147 & 0.038 & A\\
2103.61244 & 10.821 & 0.042 & A & 3144.75345 & 10.690 & 0.033 &  A & 4331.53698 & 11.076 & 0.028 & A\\
2109.60174 & 10.861 & 0.037 & A & 3152.86228 & 10.725 & 0.035 &  A & 4333.53333 & 11.140 & 0.032 & A\\
2115.57163 & 10.786 & 0.037 & A & 3154.70184 & 10.718 & 0.032 &  A & 4338.51434 & 11.133 & 0.030 & A\\
2116.57664 & 10.815 & 0.037 & A & 3162.70343 & 10.723 & 0.032 &  A & 4340.57579 & 11.176 & 0.033 & A\\
2117.57550 & 10.792 & 0.039 & A & 3164.73369 & 10.732 & 0.033 &  A & 4342.58047 & 11.172 & 0.033 & A\\
2124.55146 & 10.778 & 0.043 & A & 3167.75173 & 10.733 & 0.031 &  A & 4344.57746 & 11.205 & 0.035 & A\\
2125.56730 & 10.773 & 0.042 & A & 3169.72032 & 10.767 & 0.032 &  A & 4346.57994 & 11.239 & 0.033 & A\\
2130.56789 & 10.721 & 0.037 & A & 3171.85324 & 10.795 & 0.055 &  B & 4348.58159 & 11.206 & 0.032 & A\\
2133.55142 & 10.696 & 0.040 & A & 3178.66149 & 10.813 & 0.037 &  A & 4350.57676 & 11.231 & 0.035 & A\\
2134.54494 & 10.683 & 0.040 & A & 3180.76054 & 10.868 & 0.037 &  A & 4352.58899 & 11.300 & 0.030 & A\\
2135.54601 & 10.678 & 0.039 & A & 3183.74281 & 10.815 & 0.034 &  A & 4355.57806 & 11.267 & 0.033 & A\\
2140.51371 & 10.639 & 0.047 & B & 3191.74947 & 10.867 & 0.040 &  A & 4357.58803 & 11.319 & 0.033 & A\\
2140.52774 & 10.640 & 0.043 & A & 3266.53629 & 10.796 & 0.029 &  A & 4359.55213 & 11.276 & 0.032 & A\\
2142.51725 & 10.644 & 0.039 & A & 3278.60864 & 10.809 & 0.040 &  A & 4365.49796 & 11.282 & 0.031 & A\\
2151.49387 & 10.622 & 0.043 & A & 3292.53450 & 10.941 & 0.037 &  A & 4368.57314 & 11.309 & 0.035 & A\\
2151.51481 & 10.661 & 0.045 & A & 3298.50418 & 10.785 & 0.032 &  A & 4372.59864 & 11.345 & 0.038 & A\\
2156.48797 & 10.661 & 0.034 & A & 3404.88403 & 10.832 & 0.040 &  A & 4375.60334 & 11.429 & 0.032 & A\\
2164.47852 & 10.774 & 0.036 & A & 3414.88178 & 10.847 & 0.036 &  A & 4378.50501 & 11.420 & 0.036 & A\\
2168.54968 & 10.788 & 0.038 & A & 3418.87955 & 10.854 & 0.056 &  D & 4380.56009 & 11.418 & 0.035 & A\\
2172.48929 & 10.823 & 0.035 & A & 3422.86890 & 10.816 & 0.043 &  A & 4383.53295 & 11.538 & 0.029 & A\\
2178.48638 & 10.873 & 0.038 & A & 3426.86242 & 10.784 & 0.040 &  A & 4386.51811 & 11.697 & 0.030 & A\\
2180.53133 & 10.864 & 0.037 & A & 3446.88537 & 10.785 & 0.033 &  A & 4392.49714 & 11.891 & 0.035 & A\\
2184.53135 & 10.818 & 0.047 & B & 3450.85023 & 10.782 & 0.037 &  A & 4400.50462 & 12.333 & 0.038 & A\\
2188.51177 & 10.840 & 0.035 & A & 3453.82351 & 10.762 & 0.034 &  A & 4409.51408 & 13.217 & 0.036 & A\\
2192.50133 & 10.845 & 0.035 & A & 3456.84081 & 10.805 & 0.038 &  A & 4508.88491 & 13.357 & 0.038 & A\\
2441.66561 & 10.811 & 0.034 & A & 3463.81654 & 10.815 & 0.040 &  A & 4518.87474 & 12.488 & 0.037 & A\\
2443.65941 & 10.807 & 0.026 & A & 3469.84212 & 10.734 & 0.038 &  A & 4522.87709 & 12.563 & 0.031 & A\\
2444.65222 & 10.813 & 0.029 & A & 3472.86391 & 10.762 & 0.035 &  A & 4529.85835 & 12.504 & 0.031 & A\\
2446.66627 & 10.813 & 0.029 & A & 3475.82595 & 10.797 & 0.047 &  A & 4533.87840 & 12.534 & 0.037 & A\\
2452.62571 & 10.853 & 0.040 & A & 3478.78756 & 10.747 & 0.041 &  A & 4537.84370 & 12.662 & 0.032 & A\\
2459.61282 & 10.823 & 0.046 & A & 3481.78940 & 10.807 & 0.042 &  A & 4540.86788 & 12.460 & 0.038 & A\\
2463.61786 & 10.780 & 0.028 & A & 3483.82058 & 10.790 & 0.037 &  A & 4547.83524 & 12.400 & 0.035 & A\\
2464.61902 & 10.795 & 0.029 & A & 3489.86652 & 10.797 & 0.039 &  A & 4551.83969 & 12.399 & 0.041 & A\\
2465.62043 & 10.770 & 0.029 & A & 3491.88992 & 10.790 & 0.049 &  B & 4559.85123 & 12.105 & 0.030 & A\\
2466.60753 & 10.775 & 0.028 & A & 3499.78427 & 10.758 & 0.048 &  A & 4562.86230 & 12.096 & 0.034 & A\\
2467.61244 & 10.793 & 0.034 & A & 3502.75037 & 10.786 & 0.042 &  A & 4565.87866 & 11.901 & 0.032 & A\\
2470.53720 & 10.775 & 0.031 & A & 3504.78985 & 10.784 & 0.041 &  A & 4568.84687 & 11.909 & 0.039 & A\\
2474.62250 & 10.750 & 0.034 & A & 3510.82468 & 10.750 & 0.032 &  A & 4571.80941 & 11.698 & 0.032 & A\\
2482.59589 & 10.820 & 0.038 & A & 3517.77501 & 10.684 & 0.040 &  A & 4574.77250 & 11.621 & 0.033 & A\\
2486.57132 & 10.817 & 0.045 & B & 3521.73467 & 10.762 & 0.042 &  A & 4576.83365 & 11.570 & 0.039 & A\\
2489.56035 & 10.835 & 0.037 & A & 3523.77132 & 10.770 & 0.056 &  D & 4586.86275 & 11.342 & 0.036 & A\\
2490.57401 & 10.841 & 0.037 & A & 3525.83400 & 10.765 & 0.046 &  B & 4586.86848 & 11.311 & 0.032 & A\\
2493.56026 & 10.820 & 0.040 & A & 3528.68716 & 10.759 & 0.041 &  A & 4589.80061 & 11.282 & 0.033 & A\\
2495.56736 & 10.841 & 0.033 & A & 3530.72938 & 10.732 & 0.040 &  A & 4592.76977 & 11.270 & 0.033 & A\\
2497.55787 & 10.835 & 0.031 & A & 3539.65622 & 10.650 & 0.052 &  D & 4595.77096 & 11.207 & 0.032 & A\\
2498.54597 & 10.826 & 0.033 & A & 3544.70032 & 10.688 & 0.047 &  D & 4597.85833 & 11.194 & 0.031 & A\\
2499.53951 & 10.832 & 0.027 & A & 3547.75284 & 10.714 & 0.039 &  A & 4602.77582 & 11.205 & 0.034 & A\\
2501.54772 & 10.816 & 0.159 & D & 3551.74481 & 10.763 & 0.049 &  B & 4606.77977 & 11.262 & 0.040 & A\\
2508.57056 & 10.829 & 0.032 & A & 3553.86272 & 10.720 & 0.059 &  D & 4610.78142 & 11.173 & 0.033 & A\\
2510.53066 & 10.854 & 0.036 & A & 3556.61734 & 10.768 & 0.035 &  A & 4612.78279 & 11.175 & 0.053 & B\\
2511.59233 & 10.849 & 0.037 & A & 3559.65093 & 10.739 & 0.049 &  D & 4622.73751 & 11.175 & 0.033 & A\\
2512.65139 & 10.869 & 0.048 & A & 3561.66521 & 10.819 & 0.040 &  A & 4627.71547 & 11.166 & 0.033 & A\\
2519.65366 & 10.807 & 0.033 & A & 3563.63391 & 10.798 & 0.038 &  A & 4629.70080 & 11.244 & 0.032 & A\\
2521.65567 & 10.793 & 0.033 & A & 3574.78547 & 10.823 & 0.036 &  A & 4631.71467 & 11.226 & 0.032 & A\\
2524.61338 & 10.771 & 0.043 & B & 3576.78662 & 10.833 & 0.038 &  A & 4633.69397 & 11.213 & 0.037 & A\\
2543.51237 & 10.623 & 0.036 & A & 3584.75114 & 10.811 & 0.045 &  D & 4640.65753 & 11.151 & 0.034 & A\\
2544.49166 & 10.619 & 0.035 & A & 3593.55661 & 10.793 & 0.034 &  A & 4642.68813 & 11.155 & 0.030 & A\\
2545.55285 & 10.639 & 0.034 & A & 3599.59462 & 10.805 & 0.032 &  A & 4644.71063 & 11.135 & 0.033 & A\\
2547.53527 & 10.612 & 0.034 & A & 3601.66452 & 10.742 & 0.053 &  B & 4646.71972 & 11.132 & 0.041 & A\\
2552.52426 & 10.634 & 0.044 & B & 3603.66385 & 10.814 & 0.037 &  A & 4648.76300 & 11.126 & 0.036 & A\\
2560.48978 & 10.658 & 0.033 & A & 3606.66320 & 10.872 & 0.041 &  A & 4650.82420 & 11.068 & 0.036 & A\\
2561.51828 & 10.679 & 0.037 & A & 3616.49767 & 10.837 & 0.039 &  A & 4653.60971 & 11.124 & 0.030 & A\\
2564.52232 & 10.692 & 0.035 & A & 3618.60179 & 10.876 & 0.043 &  A & 4655.64585 & 11.067 & 0.031 & A\\
2566.49718 & 10.689 & 0.036 & A & 3620.60528 & 10.882 & 0.046 &  D & 4657.68244 & 11.105 & 0.039 & A\\
2568.53853 & 10.718 & 0.036 & A & 3628.49132 & 10.884 & 0.036 &  A & 4660.63377 & 11.058 & 0.048 & A\\
2678.87312 & 10.825 & 0.036 & A & 3630.57205 & 10.822 & 0.037 &  A & 4665.63048 & 11.051 & 0.034 & A\\
2685.86873 & 10.888 & 0.033 & A & 3632.57919 & 10.889 & 0.037 &  A & 4670.72256 & 11.035 & 0.053 & B\\
2688.87290 & 10.861 & 0.039 & A & 3634.60725 & 10.818 & 0.037 &  A & 4672.73512 & 11.041 & 0.032 & A\\
2693.86357 & 10.886 & 0.034 & A & 3637.61923 & 10.838 & 0.036 &  A & 4678.74256 & 10.958 & 0.058 & D\\
2698.85680 & 10.849 & 0.037 & A & 3641.57576 & 10.806 & 0.034 &  A & 4681.66501 & 10.919 & 0.036 & A\\
2701.84353 & 10.839 & 0.033 & A & 3643.62026 & 10.803 & 0.034 &  A & 4683.66433 & 10.854 & 0.040 & A\\
2704.84566 & 10.836 & 0.033 & A & 3647.55667 & 10.828 & 0.030 &  A & 4685.64790 & 10.896 & 0.038 & A\\
2706.86653 & 10.808 & 0.033 & A & 3654.49371 & 10.820 & 0.044 &  A & 4687.60767 & 10.875 & 0.035 & C\\
2711.83090 & 10.847 & 0.030 & A & 3659.52089 & 10.828 & 0.037 &  A & 4693.54540 & 10.885 & 0.034 & A\\
2713.86742 & 10.835 & 0.032 & A & 3661.53924 & 10.882 & 0.040 &  A & 4701.55681 & 10.923 & 0.032 & A\\
2717.84758 & 10.840 & 0.035 & A & 3663.54306 & 10.859 & 0.038 &  A & 4703.56097 & 10.931 & 0.037 & A\\
2720.81780 & 10.831 & 0.037 & A & 3666.54497 & 10.840 & 0.040 &  A & 4705.59719 & 10.898 & 0.038 & A\\
2725.78682 & 10.810 & 0.034 & A & 3670.50303 & 10.809 & 0.036 &  A & 4707.68375 & 10.941 & 0.042 & A\\
2727.84169 & 10.821 & 0.034 & A & 3676.50738 & 10.824 & 0.040 &  A & 4710.56786 & 10.897 & 0.048 & A\\
2729.83267 & 10.796 & 0.036 & A & 3767.87772 & 10.843 & 0.043 &  A & 4720.52758 & 10.854 & 0.036 & A\\
2732.86927 & 10.748 & 0.034 & A & 3774.88655 & 10.818 & 0.044 &  A & 4722.57958 & 10.850 & 0.041 & A\\
2734.83991 & 10.760 & 0.032 & A & 3791.87572 & 10.897 & 0.043 &  A & 4725.60196 & 10.869 & 0.044 & A\\
2736.81316 & 10.756 & 0.031 & A & 3795.87709 & 10.900 & 0.037 &  A & 4728.55975 & 10.872 & 0.039 & A\\
2738.82916 & 10.760 & 0.033 & A & 3803.86526 & 10.858 & 0.048 &  A & 4730.60237 & 10.881 & 0.042 & B\\
2740.80750 & 10.767 & 0.038 & A & 3806.87455 & 10.793 & 0.040 &  A & 4733.61441 & 10.901 & 0.038 & A\\
2744.79595 & 10.773 & 0.051 & B & 3810.86534 & 10.813 & 0.043 &  A & 4737.59596 & 10.926 & 0.031 & A\\
2751.86773 & 10.768 & 0.035 & A & 3814.85063 & 10.791 & 0.045 &  A & 4740.58241 & 10.954 & 0.031 & A\\
2754.80976 & 10.775 & 0.033 & A & 3817.79346 & 11.113 & 0.045 &  A & 4747.51394 & 10.886 & 0.046 & B\\
2756.77009 & 10.763 & 0.031 & A & 3817.82384 & 10.782 & 0.043 &  A & 4755.56689 & 10.922 & 0.039 & A\\
2758.80878 & 10.789 & 0.029 & A & 3817.85291 & 10.835 & 0.040 &  A & 4758.53982 & 10.915 & 0.039 & A\\
2760.75837 & 10.825 & 0.034 & A & 3817.88154 & 10.945 & 0.043 &  A & 4761.52917 & 10.895 & 0.038 & A\\
2764.74750 & 10.807 & 0.033 & A & 3817.90940 & 10.795 & 0.042 &  A & 4764.52606 & 10.885 & 0.037 & A\\
2768.83207 & 10.883 & 0.035 & A & 3818.78028 & 10.816 & 0.043 &  A & 4767.52218 & 10.871 & 0.041 & A\\
2770.79462 & 10.835 & 0.042 & A & 3818.81001 & 10.798 & 0.048 &  D & 4774.49989 & 10.901 & 0.043 & B\\
2775.79778 & 10.870 & 0.036 & A & 3818.83776 & 10.840 & 0.046 &  A & 4873.88154 & 10.899 & 0.096 & D\\
2782.77636 & 10.866 & 0.047 & B & 3818.86603 & 10.787 & 0.054 &  B & 4884.86344 & 10.890 & 0.043 & A\\
2784.73145 & 10.847 & 0.030 & A & 3818.89487 & 10.817 & 0.041 &  A & 4888.85412 & 10.844 & 0.060 & D\\
2786.69223 & 10.823 & 0.034 & A & 3819.83248 & 10.806 & 0.047 &  A & 4892.85329 & 10.871 & 0.048 & A\\
2787.85465 & 10.932 & 0.039 & A & 3822.86444 & 10.843 & 0.039 &  A & 4904.90499 & 10.908 & 0.056 & D\\
2791.65946 & 10.836 & 0.035 & A & 3825.85961 & 10.778 & 0.045 &  A & 4916.83957 & 10.879 & 0.045 & A\\
2795.74352 & 10.816 & 0.034 & A & 3829.81893 & 10.790 & 0.037 &  A & 4919.88115 & 10.824 & 0.049 & B\\
2808.70976 & 10.829 & 0.031 & A & 3832.83783 & 10.785 & 0.043 &  A & 4924.90464 & 10.796 & 0.056 & D\\
2810.63295 & 10.812 & 0.031 & A & 3835.83326 & 10.867 & 0.039 &  A & 4928.89062 & 10.813 & 0.054 & B\\
2811.81572 & 10.830 & 0.035 & A & 3847.85540 & 10.886 & 0.047 &  A & 4931.85646 & 10.816 & 0.054 & B\\
2813.70912 & 10.810 & 0.034 & A & 3850.79804 & 10.911 & 0.047 &  D & 4940.81812 & 10.797 & 0.050 & B\\
2820.70452 & 10.832 & 0.029 & A & 3852.84525 & 10.906 & 0.044 &  A & 4946.79831 & 10.859 & 0.049 & A\\
2826.73553 & 10.836 & 0.028 & A & 3855.78777 & 10.864 & 0.056 &  D & 4949.79831 & 10.840 & 0.050 & B\\
2830.61263 & 10.763 & 0.041 & A & 3858.77707 & 10.844 & 0.045 &  A & 4952.77460 & 10.835 & 0.045 & A\\
2835.62683 & 10.810 & 0.030 & A & 3860.84328 & 10.819 & 0.035 &  A & 4954.82780 & 10.803 & 0.045 & A\\
2838.62847 & 10.819 & 0.029 & A & 3862.83173 & 10.814 & 0.036 &  A & 4959.81795 & 10.814 & 0.057 & D\\
2840.68315 & 10.802 & 0.031 & A & 3864.75915 & 10.772 & 0.047 &  D & 4965.81098 & 10.813 & 0.059 & D\\
2842.66518 & 10.780 & 0.094 & D & 3866.76219 & 10.834 & 0.041 &  A & 4967.83882 & 10.826 & 0.047 & A\\
2845.67292 & 10.810 & 0.030 & A & 3868.78516 & 10.789 & 0.043 &  A & 4973.81523 & 10.862 & 0.054 & D\\
2851.58670 & 10.791 & 0.027 & A & 3872.79988 & 10.803 & 0.041 &  A & 4984.74961 & 10.855 & 0.054 & B\\
2853.71459 & 10.811 & 0.030 & A & 3882.71417 & 10.851 & 0.038 &  A & 4988.73132 & 10.868 & 0.053 & B\\
2855.62088 & 10.858 & 0.033 & A & 3892.72766 & 10.912 & 0.036 &  A & 4992.79084 & 10.808 & 0.064 & D\\
2861.56826 & 10.856 & 0.030 & A & 3894.71976 & 10.902 & 0.031 &  A & 5002.69736 & 10.862 & 0.049 & A\\
2862.67814 & 10.841 & 0.034 & A & 3900.72629 & 10.959 & 0.039 &  A & 5006.70044 & 10.927 & 0.052 & B\\
2866.67894 & 10.854 & 0.034 & A & 3902.73348 & 10.952 & 0.036 &  A & 5009.66873 & 10.899 & 0.057 & D\\
2867.73639 & 10.867 & 0.034 & A & 3904.76183 & 10.964 & 0.040 &  A & 5014.66445 & 10.927 & 0.045 & A\\
2872.52558 & 10.884 & 0.029 & A & 3906.76419 & 10.913 & 0.040 &  A & 5019.76806 & 10.889 & 0.059 & D\\
2874.72286 & 10.896 & 0.036 & A & 3908.83021 & 10.918 & 0.040 &  A & 5021.81148 & 10.934 & 0.053 & B\\
2876.64020 & 10.894 & 0.035 & A & 4152.88631 & 11.043 & 0.036 &  A & 5023.81240 & 10.899 & 0.045 & A\\
2878.63041 & 10.880 & 0.038 & A & 4152.89398 & 10.999 & 0.032 &  A & 5029.62698 & 10.887 & 0.048 & A\\
2883.55984 & 10.879 & 0.031 & A & 4156.88158 & 10.990 & 0.045 &  A & 5035.60730 & 10.924 & 0.059 & D\\
2888.49522 & 10.890 & 0.031 & A & 4164.86027 & 11.018 & 0.045 &  A & 5038.59759 & 10.942 & 0.051 & B\\
2892.49597 & 10.823 & 0.037 & A & 4168.86280 & 10.984 & 0.035 &  A & 5040.59756 & 10.908 & 0.064 & D\\
2897.61694 & 10.863 & 0.037 & A & 4174.85056 & 10.896 & 0.031 &  A & 5042.58871 & 10.893 & 0.051 & B\\
2899.60138 & 10.888 & 0.045 & A & 4177.87410 & 10.862 & 0.035 &  A & 5047.58812 & 10.883 & 0.053 & B\\
2901.59960 & 10.819 & 0.072 & D & 4181.83864 & 10.920 & 0.048 &  A & 5048.75604 & 10.831 & 0.047 & A\\
2904.64002 & 10.831 & 0.044 & A & 4184.88518 & 10.965 & 0.055 &  B & 5067.55502 & 10.835 & 0.046 & A\\
2910.50320 & 10.796 & 0.034 & A & 4188.87881 & 10.914 & 0.037 &  A & 5069.56605 & 10.858 & 0.061 & D\\
2916.49019 & 10.798 & 0.041 & A & 4191.83391 & 10.870 & 0.032 &  A & 5079.51252 & 10.865 & 0.060 & D\\
2917.59014 & 10.785 & 0.034 & A & 4194.81112 & 10.868 & 0.034 &  A & 5084.61629 & 10.867 & 0.048 & A\\
2921.58927 & 10.767 & 0.038 & A & 4202.84995 & 10.953 & 0.035 &  A & 5086.62941 & 10.904 & 0.056 & D\\
2923.57980 & 10.780 & 0.040 & A & 4205.80904 & 10.898 & 0.033 &  A & 5089.54390 & 10.889 & 0.053 & B\\
2928.57731 & 10.704 & 0.035 & A & 4213.85393 & 10.850 & 0.034 &  A & 5093.60049 & 10.963 & 0.052 & B\\
2931.51275 & 10.708 & 0.034 & A & 4216.85165 & 10.865 & 0.035 &  A & 5101.60119 & 10.906 & 0.062 & D\\
2933.52211 & 10.673 & 0.035 & A & 4228.82531 & 10.873 & 0.030 &  A & 5104.59367 & 10.885 & 0.049 & A\\
2935.52339 & 10.698 & 0.037 & A & 4230.80979 & 10.903 & 0.034 &  A & 5107.55495 & 10.854 & 0.055 & B\\
2937.52244 & 10.695 & 0.036 & A & 4232.81292 & 10.914 & 0.032 &  A & 5116.52201 & 10.726 & 0.054 & B\\
2943.50283 & 10.668 & 0.037 & A & 4234.85070 & 10.868 & 0.033 &  A & 5119.54271 & 10.731 & 0.079 & D\\
2947.50711 & 10.678 & 0.039 & A & 4245.72014 & 10.896 & 0.048 &  A & 5123.54453 & 10.748 & 0.055 & B\\
2951.50800 & 10.681 & 0.038 & A & 4247.70513 & 10.929 & 0.030 &  A & 5129.50166 & 10.733 & 0.058 & D\\
3054.88419 & 10.845 & 0.032 & A & 4250.82986 & 10.908 & 0.033 &  A & 5131.54720 & 10.771 & 0.056 & D\\
3055.81717 & 10.846 & 0.042 & A & 4255.75447 & 10.932 & 0.035 &  A & 5136.50328 & 10.794 & 0.065 & D\\
3057.82422 & 10.855 & 0.041 & A & 4258.68006 & 10.928 & 0.031 &  A & 5145.50549 & 10.754 & 0.060 & D\\
\end{longtable}
%\end{landscape}
}% End \small
}% End \longtabL

\addtocounter{table}{2}
\longtab{2}{
\small{
%\begin{landscape}
\begin{longtable}{lrlllrlllrll}
   \caption{\label{ASAS_I.dat} Photometric data obtained from the ASAS-3
survey in $I$ band.  The column labeled ``Grade" shows the quality flags,
where A denotes the highest quality of measurements and D the lowest.}\\
\hline\hline
$HJD-2450000$ & $I$ & Err$_I$ & Grade & $HJD-2450000$ & $I$ & Err$_I$ & Grade & $HJD-2450000$ & $I$ & Err$_I$ & Grade\\
\hline
\endfirsthead
\caption{continued.}\\
\hline\hline
$HJD-2450000$ & $I$ & Err$_I$ & Grade & $HJD-2450000$ & $I$ & Err$_I$ & Grade & $HJD-2450000$ & $I$ & Err$_I$ & Grade\\
\hline
\endhead
\hline
\endfoot
2404.67709 & 9.683 & 0.085 & A & 3529.72755 &  9.332 & 0.088 & D & 4532.85709 &  9.951 & 0.100 & D\\
2405.73544 & 9.665 & 0.081 & A & 3530.72771 &  9.385 & 0.069 & A & 4535.83801 & 10.006 & 0.070 & A\\
2406.74813 & 9.683 & 0.080 & A & 3537.69214 &  9.429 & 0.079 & D & 4550.85713 &  9.958 & 0.070 & A\\
2416.80251 & 9.766 & 0.074 & A & 3544.67200 &  9.407 & 0.083 & D & 4556.79372 & 10.017 & 0.071 & A\\
2679.86932 & 9.864 & 0.067 & A & 3550.67958 &  9.407 & 0.071 & A & 4560.83775 & 10.037 & 0.077 & A\\
2683.86604 & 9.834 & 0.060 & A & 3559.78678 &  9.465 & 0.072 & A & 4562.84532 & 10.038 & 0.069 & A\\
2689.86412 & 9.834 & 0.059 & A & 3563.65130 &  9.534 & 0.081 & D & 4564.81732 & 10.042 & 0.081 & A\\
2693.86237 & 9.817 & 0.060 & A & 3641.49674 &  9.771 & 0.064 & A & 4566.80186 & 10.055 & 0.071 & A\\
2701.86047 & 9.786 & 0.058 & A & 3643.51692 &  9.769 & 0.071 & A & 4568.80495 & 10.062 & 0.079 & A\\
2707.84277 & 9.787 & 0.059 & A & 3646.53610 &  9.788 & 0.074 & A & 4570.79733 & 10.042 & 0.074 & A\\
2713.83075 & 9.767 & 0.062 & A & 3654.49360 &  9.783 & 0.071 & A & 4574.78885 & 10.027 & 0.075 & A\\
2720.81842 & 9.758 & 0.061 & A & 3656.53296 &  9.815 & 0.069 & A & 4576.78802 & 10.024 & 0.081 & A\\
2725.78561 & 9.694 & 0.056 & A & 3658.57870 &  9.780 & 0.067 & A & 4583.85374 & 10.055 & 0.068 & A\\
2734.90368 & 9.649 & 0.056 & A & 3661.50212 &  9.784 & 0.066 & A & 4588.76459 & 10.024 & 0.068 & A\\
2739.81681 & 9.619 & 0.050 & A & 3663.51042 &  9.806 & 0.081 & A & 4590.75356 & 10.046 & 0.078 & A\\
2741.82374 & 9.643 & 0.123 & D & 3676.49937 &  9.752 & 0.069 & A & 4592.74225 & 10.057 & 0.074 & A\\
2747.82485 & 9.623 & 0.062 & A & 3681.50133 &  9.719 & 0.071 & A & 4594.74559 & 10.044 & 0.080 & A\\
2755.91076 & 9.593 & 0.057 & A & 3766.88239 &  9.661 & 0.084 & A & 4595.82503 & 10.061 & 0.081 & A\\
2759.83233 & 9.569 & 0.059 & A & 3772.88718 &  9.641 & 0.080 & A & 4597.74851 & 10.076 & 0.080 & A\\
2761.77879 & 9.576 & 0.056 & A & 3775.88633 &  9.706 & 0.070 & A & 4599.72779 & 10.082 & 0.072 & A\\
2763.78161 & 9.573 & 0.064 & A & 3778.87774 &  9.691 & 0.069 & A & 4601.72516 & 10.084 & 0.072 & A\\
2765.78437 & 9.626 & 0.064 & A & 3781.86909 &  9.711 & 0.070 & A & 4603.71056 & 10.115 & 0.080 & A\\
2767.74827 & 9.596 & 0.063 & A & 3814.84161 &  9.766 & 0.075 & A & 4610.78129 & 10.147 & 0.067 & A\\
2770.75197 & 9.638 & 0.063 & A & 3819.82327 &  9.790 & 0.078 & A & 4612.70961 & 10.139 & 0.069 & A\\
2775.79622 & 9.620 & 0.070 & A & 3822.85359 &  9.802 & 0.076 & A & 4616.76163 & 10.135 & 0.077 & A\\
2782.77742 & 9.648 & 0.065 & A & 3825.81839 &  9.774 & 0.075 & A & 4618.74702 & 10.145 & 0.096 & D\\
2826.66055 & 9.800 & 0.095 & D & 3828.85882 &  9.774 & 0.070 & A & 4623.67579 & 10.178 & 0.073 & A\\
2844.75061 & 9.796 & 0.076 & A & 3831.79541 &  9.749 & 0.071 & A & 4627.66019 & 10.181 & 0.074 & A\\
2851.58802 & 9.784 & 0.070 & A & 3833.81871 &  9.771 & 0.074 & A & 4628.78570 & 10.164 & 0.092 & D\\
2861.56822 & 9.815 & 0.068 & A & 3837.81014 &  9.852 & 0.075 & A & 4631.72978 & 10.112 & 0.070 & A\\
2862.67908 & 9.836 & 0.082 & A & 3852.85189 &  9.787 & 0.074 & A & 4632.86368 & 10.141 & 0.071 & A\\
2866.59195 & 9.831 & 0.081 & A & 3856.83575 &  9.807 & 0.075 & A & 4637.78642 & 10.141 & 0.084 & A\\
2867.70877 & 9.825 & 0.090 & D & 3858.83984 &  9.761 & 0.079 & A & 4639.71546 & 10.104 & 0.072 & A\\
2871.70949 & 9.839 & 0.066 & A & 3862.74402 &  9.777 & 0.074 & A & 4640.85128 & 10.127 & 0.075 & A\\
2874.65233 & 9.812 & 0.086 & A & 3864.74260 &  9.799 & 0.081 & A & 4645.72970 & 10.107 & 0.073 & A\\
2878.63296 & 9.774 & 0.075 & A & 3866.73119 &  9.791 & 0.076 & A & 4650.61704 & 10.128 & 0.071 & A\\
2880.68999 & 9.762 & 0.082 & A & 3872.80064 &  9.789 & 0.097 & D & 4651.78255 & 10.102 & 0.072 & A\\
2883.53073 & 9.786 & 0.074 & A & 3900.72872 &  9.659 & 0.072 & A & 4660.60399 & 10.015 & 0.068 & A\\
2888.49366 & 9.734 & 0.082 & A & 3902.65477 &  9.668 & 0.077 & A & 4665.63142 &  9.982 & 0.069 & A\\
2892.49347 & 9.723 & 0.087 & A & 3905.73613 &  9.652 & 0.081 & A & 4666.78240 &  9.952 & 0.070 & A\\
2893.60679 & 9.719 & 0.071 & A & 3913.74822 &  9.565 & 0.070 & A & 4670.56897 &  9.964 & 0.075 & A\\
2897.58470 & 9.694 & 0.069 & A & 3919.82413 &  9.536 & 0.084 & A & 4675.73254 &  9.970 & 0.097 & D\\
2899.57886 & 9.701 & 0.072 & A & 3921.81330 &  9.536 & 0.084 & A & 4684.60377 &  9.856 & 0.071 & A\\
3402.88317 & 9.748 & 0.072 & A & 3933.66495 &  9.515 & 0.077 & A & 4693.52078 &  9.803 & 0.067 & A\\
3409.88339 & 9.758 & 0.069 & A & 3935.67622 &  9.491 & 0.076 & A & 4694.64085 &  9.813 & 0.073 & A\\
3413.88165 & 9.773 & 0.073 & A & 3937.67151 &  9.459 & 0.073 & A & 4697.71364 &  9.844 & 0.078 & A\\
3417.87584 & 9.762 & 0.067 & A & 3940.60651 &  9.464 & 0.078 & A & 4701.63585 &  9.852 & 0.071 & A\\
3421.87626 & 9.769 & 0.069 & A & 3944.60782 &  9.458 & 0.081 & A & 4720.52715 &  9.703 & 0.068 & A\\
3424.87269 & 9.745 & 0.070 & A & 3946.60328 &  9.488 & 0.076 & A & 4739.52992 &  9.757 & 0.067 & A\\
3427.85797 & 9.739 & 0.072 & A & 3950.60339 &  9.474 & 0.064 & A & 4747.49318 &  9.754 & 0.069 & A\\
3430.87108 & 9.782 & 0.069 & A & 3957.60129 &  9.529 & 0.068 & A & 4753.57172 &  9.733 & 0.067 & A\\
3443.87383 & 9.754 & 0.065 & A & 4177.83332 &  9.653 & 0.100 & D & 4766.52246 &  9.756 & 0.070 & A\\
3446.85333 & 9.759 & 0.065 & A & 4180.82502 &  9.706 & 0.092 & D & 4768.52121 &  9.783 & 0.071 & A\\
3456.83873 & 9.692 & 0.066 & A & 4182.89967 &  9.663 & 0.085 & A & 4774.49956 &  9.817 & 0.069 & A\\
3463.81395 & 9.690 & 0.074 & A & 4184.85894 &  9.712 & 0.086 & A & 4778.50690 &  9.803 & 0.069 & A\\
3465.84898 & 9.685 & 0.065 & A & 4190.83881 &  9.715 & 0.115 & D & 4862.87803 &  9.972 & 0.076 & A\\
3467.87378 & 9.672 & 0.065 & A & 4192.82234 &  9.737 & 0.093 & D & 4866.88608 &  9.934 & 0.073 & A\\
3473.83732 & 9.624 & 0.065 & A & 4194.82809 &  9.762 & 0.099 & D & 4878.86047 &  9.874 & 0.073 & A\\
3475.87195 & 9.648 & 0.065 & A & 4202.81954 &  9.821 & 0.100 & D & 4884.85659 &  9.844 & 0.067 & A\\
3477.80831 & 9.658 & 0.067 & A & 4204.83177 &  9.772 & 0.093 & D & 4891.85643 &  9.782 & 0.073 & A\\
3479.79468 & 9.634 & 0.072 & A & 4207.83850 &  9.808 & 0.100 & D & 4897.87710 &  9.744 & 0.070 & A\\
3481.79055 & 9.618 & 0.071 & A & 4213.78094 &  9.832 & 0.093 & D & 4901.85042 &  9.744 & 0.072 & A\\
3483.78151 & 9.640 & 0.075 & A & 4216.78003 &  9.851 & 0.115 & D & 4906.85356 &  9.726 & 0.068 & A\\
3489.86792 & 9.587 & 0.078 & D & 4228.82693 &  9.855 & 0.111 & D & 4911.80876 &  9.725 & 0.070 & A\\
3491.79078 & 9.616 & 0.069 & A & 4230.73180 &  9.865 & 0.103 & D & 4913.87981 &  9.722 & 0.070 & A\\
3495.79351 & 9.605 & 0.073 & A & 4232.74030 &  9.863 & 0.103 & D & 4917.82412 &  9.691 & 0.073 & A\\
3497.82332 & 9.597 & 0.070 & A & 4234.74033 &  9.860 & 0.094 & D & 4920.82387 &  9.642 & 0.069 & A\\
3500.80823 & 9.562 & 0.064 & A & 4238.76823 &  9.840 & 0.103 & D & 4924.86424 &  9.672 & 0.074 & A\\
3504.76931 & 9.539 & 0.066 & A & 4240.72245 &  9.867 & 0.096 & D & 4926.83146 &  9.632 & 0.071 & A\\
3508.78634 & 9.478 & 0.072 & A & 4243.73437 &  9.864 & 0.115 & D & 4929.77815 &  9.616 & 0.067 & A\\
3510.83457 & 9.449 & 0.071 & A & 4245.69219 &  9.887 & 0.109 & D & 4931.81923 &  9.672 & 0.075 & A\\
3517.77423 & 9.440 & 0.069 & A & 4246.76654 &  9.923 & 0.115 & D & 4933.81043 &  9.650 & 0.076 & A\\
3521.68389 & 9.416 & 0.082 & A & 4250.73531 &  9.922 & 0.108 & D & 4939.79739 &  9.628 & 0.073 & A\\
3522.94456 & 9.415 & 0.072 & A & 4494.87766 & 10.919 & 0.071 & A & 4943.79401 &  9.659 & 0.074 & A\\
3525.70579 & 9.441 & 0.067 & A & 4502.88696 & 10.619 & 0.076 & A & 4947.80194 &  9.642 & 0.071 & A\\
3527.70083 & 9.418 & 0.073 & A & 4506.87894 & 10.436 & 0.070 & A & 4951.76611 &  9.647 & 0.069 & A\\
3528.74064 & 9.429 & 0.069 & A & 4510.86941 & 10.348 & 0.071 & A & 4953.76489 &  9.638 & 0.070 & A\\
3529.67170 & 9.407 & 0.075 & A & 4513.86912 & 10.148 & 0.073 & A & 4955.75240 &  9.626 & 0.074 & A\\
3529.70102 & 9.418 & 0.086 & D & 4516.86599 & 10.106 & 0.076 & A & 4959.81727 &  9.612 & 0.081 & A\\
3529.70323 & 9.443 & 0.089 & D & 4519.85726 & 10.070 & 0.078 & A & 4965.80763 &  9.653 & 0.075 & A\\
3529.70540 & 9.346 & 0.085 & D & 4522.86223 & 10.013 & 0.077 & A & 4986.75120 &  9.773 & 0.090 & D\\
3529.72533 & 9.381 & 0.084 & D & 4529.84038 &  9.980 & 0.082 & A &            &        &       &  \\
\end{longtable}
%\end{landscape}
}% End \small
}% End \longtabL

\begin{table*}
   \caption{$V-I$ colour indices obtained from data of Tables
\ref{ASAS_V.dat} and \ref{ASAS_I.dat} when both the $V$ and $I$ measurements
on the same night were available.}
   \label{ASAS_V-I.dat}
\centering % used for centering table
\begin{tabular}{llrlrlrl}  
\hline\hline
$HJD-2450000$ & $V-I$ & $HJD-2450000$ & $V-I$ & $HJD-2450000$ & $V-I$ & $HJD-2450000$ & $V-I$ \\
\hline
2693.86297 & 1.069 & 3475.84895 & 1.149 & 3852.84857 & 1.119 & 4576.81084 & 1.546\\
2701.85200 & 1.053 & 3481.78998 & 1.189 & 3858.80846 & 1.083 & 4592.75601 & 1.213\\
2713.84909 & 1.068 & 3483.80105 & 1.150 & 3862.78788 & 1.037 & 4595.79800 & 1.146\\
2720.81811 & 1.073 & 3489.86722 & 1.210 & 3864.75088 & 0.973 & 4597.80342 & 1.118\\
2725.78622 & 1.116 & 3491.84035 & 1.174 & 3866.74669 & 1.043 & 4610.78136 & 1.026\\
2734.87180 & 1.111 & 3504.77958 & 1.245 & 3872.80026 & 1.014 & 4612.74620 & 1.036\\
2770.77330 & 1.197 & 3510.82963 & 1.301 & 3900.72751 & 1.300 & 4627.68783 & 0.985\\
2775.79700 & 1.250 & 3517.77462 & 1.244 & 3902.69413 & 1.284 & 4631.72223 & 1.114\\
2782.77689 & 1.218 & 3521.70928 & 1.346 & 4177.85371 & 1.209 & 4640.75441 & 1.024\\
2826.69804 & 1.036 & 3525.76990 & 1.324 & 4184.87206 & 1.253 & 4650.72062 & 0.940\\
2851.58736 & 1.007 & 3528.71390 & 1.330 & 4194.81961 & 1.106 & 4660.61888 & 1.043\\
2861.56824 & 1.041 & 3530.72855 & 1.347 & 4202.83475 & 1.132 & 4665.63095 & 1.069\\
2862.67861 & 1.005 & 3544.68616 & 1.281 & 4213.81744 & 1.018 & 4670.64577 & 1.071\\
2866.63545 & 1.023 & 3559.71886 & 1.274 & 4216.81584 & 1.014 & 4701.59633 & 1.071\\
2867.72258 & 1.042 & 3563.64261 & 1.264 & 4228.82612 & 1.018 & 4720.52737 & 1.151\\
2874.68760 & 1.084 & 3641.53625 & 1.035 & 4230.77080 & 1.038 & 4747.50356 & 1.132\\
2878.63169 & 1.106 & 3643.56859 & 1.034 & 4232.77661 & 1.051 & 4774.49973 & 1.084\\
2883.54529 & 1.093 & 3654.49366 & 1.037 & 4234.79552 & 1.008 & 4884.86002 & 1.046\\
2888.49444 & 1.156 & 3661.52068 & 1.098 & 4245.70617 & 1.009 & 4924.88444 & 1.124\\
2892.49472 & 1.100 & 3663.52674 & 1.053 & 4250.78259 & 0.986 & 4931.83785 & 1.144\\
2897.60082 & 1.169 & 3676.50338 & 1.072 & 4522.86966 & 2.550 & 4959.81761 & 1.202\\
2899.59012 & 1.187 & 3814.84612 & 1.025 & 4529.84937 & 2.524 & 4965.80931 & 1.160\\
3446.86935 & 1.026 & 3819.82788 & 1.016 & 4562.85381 & 2.058 &            &      \\
3456.83977 & 1.113 & 3822.85902 & 1.041 & 4568.82591 & 1.847 &            &      \\
3463.81525 & 1.125 & 3825.83900 & 1.004 & 4574.78068 & 1.594 &            &      \\
\hline
\end{tabular}
\end{table*}

\begin{table*}
   \caption{$V-I$ colour indices obtained from data of Tables
\ref{ASAS_V.dat} and \ref{ASAS_I.dat} via interpolation in the cases when
close measurements in $V$ and $I$ bands vere used while the simultaneous
measurements on the same night were not available.}
   \label{ASAS_V-I_int.dat}
\centering % used for centering table
\begin{tabular}{llrlrlrl}  
\hline\hline
$HJD-2450000$ & $V-I$ & $HJD-2450000$ & $V-I$ & $HJD-2450000$ & $V-I$ & $HJD-2450000$ & $V-I$ \\
\hline
2681.86931 & 1.008 & 3421.87626 & 1.056 & 4205.80904 & 1.114 & 4767.52218 & 1.101\\
2686.86795 & 1.041 & 3422.86890 & 1.055 & 4508.88491 & 2.965 & 4873.88154 & 1.000\\
2707.84277 & 1.029 & 3424.87269 & 1.055 & 4518.87474 & 2.406 & 4891.85643 & 1.082\\
2739.81681 & 1.144 & 3426.86242 & 1.043 & 4533.87840 & 2.564 & 4892.85329 & 1.095\\
2740.80750 & 1.136 & 3471.35301 & 1.104 & 4535.83801 & 2.591 & 4899.86376 & 1.149\\
2741.82374 & 1.126 & 3478.78756 & 1.101 & 4550.85713 & 2.441 & 4904.90499 & 1.175\\
2744.79595 & 1.140 & 3499.78427 & 1.184 & 4551.83969 & 2.431 & 4916.83957 & 1.180\\
2747.82485 & 1.148 & 3500.80823 & 1.211 & 4556.79372 & 2.200 & 4917.82412 & 1.170\\
2753.33875 & 1.169 & 3550.67958 & 1.343 & 4559.85123 & 2.073 & 4919.88115 & 1.167\\
2755.91076 & 1.175 & 3767.87772 & 1.185 & 4565.87866 & 1.852 & 4928.89062 & 1.192\\
2757.78943 & 1.201 & 3772.88718 & 1.184 & 4586.86562 & 1.291 & 4940.81812 & 1.161\\
2759.83233 & 1.239 & 3774.88655 & 1.134 & 4588.76459 & 1.274 & 4943.79401 & 1.169\\
2760.75837 & 1.253 & 3829.81893 & 1.024 & 4589.80061 & 1.247 & 4946.79831 & 1.213\\
2762.78020 & 1.241 & 3831.79541 & 1.038 & 4602.77582 & 1.105 & 4947.80194 & 1.211\\
2764.74750 & 1.208 & 3832.83783 & 1.025 & 4603.71056 & 1.103 & 4949.79831 & 1.195\\
2766.76632 & 1.234 & 3833.81871 & 1.041 & 4622.73751 & 1.003 & 4951.76611 & 1.190\\
2844.75061 & 1.005 & 3835.83326 & 1.055 & 4633.69397 & 1.072 & 4952.77460 & 1.193\\
3404.88403 & 1.081 & 3837.81014 & 1.018 & 4645.72970 & 1.026 & 4953.76489 & 1.182\\
3411.88252 & 1.076 & 3905.73613 & 1.287 & 4675.73254 & 1.030 & 4954.82780 & 1.171\\
3414.88178 & 1.077 & 4181.83864 & 1.235 & 4684.60377 & 1.018 & 4986.75120 & 1.089\\
3417.87584 & 1.090 & 4190.83881 & 1.170 & 4739.52992 & 1.187 &            &      \\
3418.87955 & 1.090 & 4191.83391 & 1.144 & 4766.52246 & 1.120 &            &      \\
\hline
\end{tabular}
\end{table*}

\end{appendix}

\end{document}